\documentclass[twocolumn]{svjour3}
\usepackage[]{graphicx}\usepackage[]{color}
\makeatletter
\def\maxwidth{ %
  \ifdim\Gin@nat@width>\linewidth
    \linewidth
  \else
    \Gin@nat@width
  \fi
}
\makeatother

\definecolor{fgcolor}{rgb}{0.345, 0.345, 0.345}

\usepackage{framed}
\makeatletter
 {\par\unskip\endMakeFramed%
 \at@end@of@kframe}
\makeatother

\definecolor{shadecolor}{rgb}{.97, .97, .97}
\definecolor{messagecolor}{rgb}{0, 0, 0}
\definecolor{warningcolor}{rgb}{1, 0, 1}
\definecolor{errorcolor}{rgb}{1, 0, 0}
\newenvironment{knitrout}{}{} 

\usepackage{alltt}
\newcommand{\SweaveOpts}[1]{}  
\newcommand{\SweaveInput}[1]{} 
\newcommand{\Sexpr}[1]{}       

\usepackage{graphicx}
\usepackage{bm}
\usepackage{amsmath}
\usepackage{dsfont}
\usepackage{booktabs}
\usepackage{algorithm}
\usepackage{csquotes}
\usepackage{subcaption}
\usepackage{hyperref}
\usepackage[utf8]{inputenc}
\usepackage[T1]{fontenc}

\DeclareMathOperator*{\argmin}{\arg\!\min} 
\newcommand{\stlen}{\textrm{sl}} 

\title{Stability selection for component-wise gradient boosting in multiple dimensions}

\author{Janek Thomas \and
        Andreas Mayr \and
        Bernd Bischl \and
        Matthias Schmid \and
        Adam Smith \and
        Benjamin Hofner}

\institute{%
    J. Thomas \and B. Bischl
        \at Department of Statistics, Ludwig-Maximilians-Universit{\"a}t M{\"u}nchen, Ludwigstrasse 33, 80539 Munich, Germany\\ Tel.: +4989-2180-3196\\
        \email{janek.thomas@stat.uni-muenchen.de} \and 
    A. Mayr 
        \at Department of Medical Informatics, Biometry
and Epidemiology, FAU Erlangen-N{\"u}rnberg, Germany \and
    M. Schmid \and A.Mayr 
        \at Department of Medical Biometry, Informatics
and Epidemiology, RFWU Bonn, Germany \and
    A. Smith
        \at U.S. Fish \& Wildlife Service, National Wildlife Refuge System, Southeast Inventory \& Monitoring Branch, USA \and
    B. Hofner 
        \at Section Biostatistics, Paul-Ehrlich-Institute, Langen, Germany
}

\begin{document}

\maketitle

\begin{abstract}

We present a new algorithm for boosting generalized additive models for location, scale and shape (GAMLSS) that allows to incorporate stability selection,
an increasingly popular way to obtain stable sets of covariates while controlling the per-family error rate (PFER). The model is fitted repeatedly to subsampled data and variables with high selection frequencies are extracted. 
To apply stability selection to boosted GAMLSS, we develop a new \enquote{noncyclical} fitting algorithm that incorporates an additional selection step of the best-fitting distribution parameter in each iteration. This new algorithms has the additional advantage that optimizing the tuning parameters of boosting is reduced from a multi-dimensional to a one-dimensional problem with vastly decreased complexity. The performance of the novel algorithm is evaluated in an extensive simulation study. We apply this new algorithm to a study to estimate abundance of common eider in Massachusetts, USA, featuring excess zeros, overdispersion, non-linearity and spatio-temporal structures. Eider abundance is estimated via boosted GAMLSS, allowing both mean and overdispersion to be regressed on covariates. Stability selection is used to obtain a sparse set of stable predictors.
\end{abstract}
\section{Introduction}
\label{sec:intro}

In view of the growing size and complexity of modern databases, statistical modeling is increasingly faced with heteroscedasticity issues and a large number of available modeling options. In ecology, for example, it is often observed that outcome variables do not only show differences in {\em mean} conditions but also tend to be highly {\em variable} across different geographical features or states of a combination of covariates (e.g., \cite{Osorio:2012}). In addition, ecological databases typically contain large numbers of correlated predictor variables that need to be carefully chosen for possible incorporation in a statistical regression model \cite{Aho:2014,Dormann,Murtaugh:2009}.

A convenient approach to address both heteroscedasticity and variable selection in statistical regression models is the combination of GAMLSS modeling with gradient boosting algorithms. GAMLSS, which refer to ``generalized additive models for location, scale and shape'' \cite{rigby2005generalized}, are a modeling technique that relates not only the mean but all parameters of the outcome distribution to the available covariates. Consequently, GAMLSS simultaneously fit different submodels for the location, scale and shape parameters of the conditional distribution. Gradient boosting, on the other hand, has become a popular tool for data-driven variable selection in generalized additive models \cite{buhlmann2007boosting}. The most important feature of gradient boosting is the ability of the algorithm to perform variable selection in each iteration, so that model fitting and variable selection are accomplished in a single algorithmic procedure. To combine GAMLSS with gradient boosting, we have developed the gamboostLSS algorithm \cite{mayr2012generalized} and have implemented this procedure in the R add-on package {\bf gamboostLSS} \cite{gamboostLSS2015tut,hofner2016gamboostLSS}.

A remaining problem of gradient boosting is the tendency of boosting algorithms to select a relatively high number of false positive variables and to include too many non-informative covariates in a statistical regression model. This issue, which has been raised in several previous articles \cite{Hothorn:Twin:2010,buhlmann2006sparse,Ma:Gen:2012}, is particularly relevant for model building in the GAMLSS framework, as the inclusion of non-informative false positives in the submodels for the scale and shape parameters may result in overfitting with a highly inflated variance. As a consequence, it is crucial to include only those covariates in these submodels that show a relevant effect on the outcome parameter of interest. From an algorithmic point of view, this problem is aggravated by the conventional fitting procedure of gamboostLSS: although the fitting procedure proposed in \cite{mayr2012generalized} incorporates different iteration numbers for each of the involved submodels, the algorithm starts with mandatory updates of each submodel at the beginning of the procedure. Consequently, due to the tendency of gradient boosting to include relatively high numbers of non-informative covariates, false positive variables may enter a GAMLSS submodel at a very early stage, even before the iteration number of the submodel is finally reached.

To address these issues and to enforce sparsity in\\ 
GAMLSS, we propose a novel procedure that incorporates {\em stability selection} \cite{Meinshausen:Stab:2010} in gamboostLSS. Stability selection is a generic method that investigates the importance of covariates in a statistical model by repeatedly subsampling the data. Sparsity is enforced by including only the most ``stable'' covariates, in the final statistical model. Importantly, under appropriate regularity conditions, stability selection can be tuned such that the expected number of false positive covariates is controlled in a mathematically rigorous way. As will be demonstrated in Section \ref{sec:simulation} of this paper, the same property holds in the gamboostLSS framework.

To combine gamboostLSS with stability selection, we present an improved ``\textit{noncyclical}'' fitting procedure for gamboostLSS that addresses the problem of possible false positive inclusions at early stages of the algorithm. In contrast to the original ``\textit{cyclical}'' gamboostLSS algorithm presented in Mayr et al. \cite{mayr2012generalized}, the new version of gamboostLSS not only performs variable selection in each iteration but additionally an iteration-wise selection of the best submodel (location, scale, or shape) that leads to the largest improvement in model fit. As a consequence, sparsity is not only enforced by the inclusion of the most ``stable'' covariates in the GAMLSS submodels but also by a data-driven choice of iteration-wise submodel updates. It is this procedure that theoretically justifies and thus enables the use of stability selection in gamboostLSS.

A further advantage of ``\textit{noncyclical}'' fitting is that the maximum number of boosting iterations for each submodel does not have to be specified individually for each submodel (as in the originally proposed ``\textit{cyclical}'' variant), instead only the overall number of iterations must be chosen optimally. Tuning the complete model reduces from a multi-dimensional to a one-dimensional optimization problem, regardless of the number of submodels, therefore drastically reducing the amount of needed runtime for model selection.

A similar approach for noncyclical fitting of\\ 
multi-parameter models was recently suggested by \cite{RePEc:inn:wpaper:2016-04} for the specific application of ensemble post-processing for weather forecasts. Our proposed method generalizes this approach, allowing gamboostLSS to be combined with stability selection in a generic way that applies to a large number of outcome distributions. 

The rest of this paper is organized as follows: In Section \ref{sec:methods} we describe the gradient boosting, GAMLSS and stability selection techniques, and show how to combine the three approaches in a single algorithm. In addition, we provide details on the new gamboostLSS fitting procedure. Results of extensive simulation studies are presented in Section \ref{sec:simulation}. They demonstrate that combining gamboostLSS with stability selection results in prediction models that are both easy to interpret and show a favorable behavior with regard to variable selection. They also show that the new gamboostLSS fitting procedure results in a large decrease in runtime while showing similar empirical convergence rates as the traditional gamboostLSS procedure. We present an application of the proposed algorithm to a spatio-temporal data set on sea duck abundance in Nantucket Sound, USA, in Section \ref{sec:application}. Section \ref{sec:conclusion} summarizes the main findings and provides details on the implementation of the proposed methodology in the R package {\bf gamboostLSS} \cite{hofner2016gamboostLSS}.

\section{Methods}
\label{sec:methods}

\subsection{Gradient boosting}
\label{sec:boosting}

\emph{Gradient boosting} is a supervised learning technique that combines an ensemble of \emph{base-learners} to estimate complex statistical dependencies. 
Base-learners should be \emph{weak} in the sense that they only possess moderate prediction accuracy, usually assumed to be at least slightly better than a random predictor, but on the other hand easy and fast to fit. Base-learners can be for example simple linear regression models, regression splines with low degrees of freedom, or stumps (i.e., trees with only one split) \cite{buhlmann2007boosting}.
One base-learner by itself will usually not be enough to fit a well performing statistical model to the data, but a boosted combination of a large number can compete with other state-of-the-art algorithms on many tasks, e.g., classification \cite{li2012robust} or image recognition \cite{opelt2004weak}.

Let $D = \{(\bm x^{(i)}, y^{(i)})\}_{i=1,...n}$ be a learning data set sampled i.i.d. from
a distribution over the joint space $\mathcal{X} \times \mathcal{Y}$, with a $p$-dimensional input space $\mathcal{X} = (\mathcal{X}_1\times \mathcal{X}_2\times...\times \mathcal{X}_p)$ and a usually one-dimensional output space $\mathcal{Y}$. The response variable is estimated through an additive model where $\mathds{E}(y | \bm x) = g^{-1}(\eta(\bm x))$, with link function $g$ and additive predictor $\eta:\mathcal{X}\rightarrow\mathds{R}$, 

\begin{equation}
\label{eq:add_pred}
\eta(\bm x) = \beta_0 + \sum^J_{j = 1}f_j(\bm x|\bm \beta_j),
\end{equation}
with a constant intercept coefficient $\beta_0$ and additive effects $f_j(\bm x | \bm \beta_j)$ derived from the pre-defined set of base-learners.  
These are usually (semi-)parametric effects, e.g., splines, with parameter vector $\bm \beta_j$.
Note that some effects may later be estimated as $0$, i.e., $f_j(\bm x|\bm \beta_j) = 0$. 
In many cases, each base-learner is defined on exactly one element $\mathcal{X}_j$ of $\mathcal{X}$ and thus Equation~\ref{eq:add_pred} simplifies to
\begin{equation}
\eta(\bm x) = \beta_0 + \sum^p_{j = 1}f_j(x_j|\bm \beta_j).
\end{equation}
To estimate the parameters $\bm\beta_1, ..., \bm\beta_J$ of the additive predictor, the boosting algorithm minimizes the \emph{empirical risk} $R$ which is the loss $\rho:\mathcal{Y} \times \mathds{R} \rightarrow \mathds{R}$ summed over all training data:
\begin{equation}
\label{eq:risk}
R = \sum^n_{i = 1}\rho(y^{(i)}, \eta(\bm x^{(i)})).
\end{equation}
The loss function measures the discrepancy between the true outcome $y^{(i)}$ and the additive predictor $\eta(\bm x^{(i)})$. Examples are the absolute loss $|y^{(i)} - \eta(\bm x^{(i)})|$, leading to a regression model for the median, the quadratic loss $(y^{(i)} - \eta(\bm x^{(i)}))^2$, leading to the conventional (mean) regression model or the binomial loss $-y^{(i)}\eta(\bm x^{(i)}) + \log(1 + \exp(\eta(\bm x^{(i)})))$ often used in classification of binary outcomes $y^{(i)} \in \{0,1\}$. Very often the loss is derived from the negative log-likelihood of the distribution of $\mathcal{Y}$, depending on the desired model \cite{friedman2000additive}. 

While there exist different types of gradient boosting algorithms \cite{mayr2014evolution,mayr2014extending}, in this article we will focus on\\
component-wise gradient boosting \cite{buehlmann03,buhlmann2007boosting}. The main idea is to fit simple regression type base-learners $h(\cdot)$ one-by-one to the negative gradient vector of the loss $\bm u = (u^{(1)},\dots,u^{(n)})$ instead of to the true outcomes $\bm y = (y^{(1)},\dots,y^{(n)})$.  Base-learners are chosen in such a way that they approximate the effect $\hat{f}(\bm x|\bm\beta_j) = \sum_m h_{j}(\cdot)$. The negative gradient vector in iteration $m$, evaluated at the estimated additive predictor $\hat{\eta}^{[m-1]}(\bm x^{(i)})$, is defined as
\begin{eqnarray*}
 \boldsymbol{u}  = \left( - \left. \frac{\partial}{\partial \eta} \rho(y, \eta) \right|_{\eta = \hat{\eta}^{[m-1]}(\bm x^{(i)}),\, y =y^{(i)}} \right)_{i = 1,...,n} \ .
\end{eqnarray*}

In every boosting iteration, each base-learner is fitted separately to the negative gradient vector by \\
least-squares or penalized least-squares regression. 
The best-fitting base-learner is selected based on the residual sum of squares with respect to $\bm {u}$
\begin{equation}
\label{eq:rss}
j^* = \argmin_{j \in 1 \ldots J}\sum^n_{i = 1}(u^{(i)} - \hat h_j(\bm x^{(i)}))^2.
\end{equation}
Only the best-performing base-learner $\hat h_{j^*}(\bm x)$ will be used to update the current additive predictor,
\begin{equation}
\label{eq:boost_one}
\hat\eta^{[m]} = \hat\eta^{[m-1]} + \stlen \cdot \hat{h}_{j^*}(\bm x)
\end{equation}  
where $0 <\stlen \ll 1$ denotes the step-length (learning rate; usually $\stlen = 0.1$). The choice of $\stlen$ is not of critical importance as long as it is sufficiently small \cite{schmid2008boosting}.

The main tuning parameter for gradient boosting algorithms is the number of iterations $m$ that are performed before the algorithm is stopped (denoted as $m_{\text{stop}}$). 
The selection of $m_{\text{stop}}$ has a crucial influence on the prediction performance of the model. 
If $m_{\text{stop}}$ is set too small, the model cannot fully incorporate the influence of the effects on the response and will consequently have a poor performance. 
On the other hand, too many iterations will result in \emph{overfitting}, which hampers the interpretation and generalizability of the model.  

\subsection{GAMLSS}

In classical generalized additive models (GAM, \cite{hastie1990generalized}) it is assumed that the conditional distribution of $\mathcal{Y}$ depends only on one parameter, usually the conditional mean. If the distribution has multiple parameters, all but one are considered to be constant nuisance parameters. 
This assumption will not always hold and should be critically examined, e.g., the assumption of constant variance is not adequate for heteroscedastic data. 
Potential dependency of the scale (and shape) parameter(s) of a distribution on predictors can be modeled in a similar way to the conditional mean (i.e., location parameter). This extended model class is called \emph{generalized additive models for location, scale and shape} (GAMLSS, \cite{rigby2005generalized}). 

The framework hence fits different prediction functions to multiple distribution parameters $\bm \theta = (\theta_1, \ldots, \theta_k), k = 1, \ldots, 4$. Given a conditional density $p(y|\bm \theta)$, one estimates additive predictors (cf.~Equation~\ref{eq:add_pred}) for each of the parameters $\theta_k$
\begin{equation}
\label{eq:multstatboost}
\eta_{\theta_k} = \beta_{0\theta_k} + \sum_{j= 1}^{J_k}f_{j\theta_k}(\bm x | \bm\beta_{j\theta_k}),\qquad k = 1, \ldots, 4. 
\end{equation}
Typically these models are estimated via penalized likelihood. 
For details on the fitting algorithm see \cite{stasinopoulos2008Instructions}. 

Even though these models can be applied to a large number of different situations, and the available fitting algorithms are extremely powerful, they still inherit some shortcomings from the penalized likelihood approach: 
\begin{enumerate}
\item[(1)] It is not possible to estimate models with more covariates than observations.
\item[(2)] Maximum likelihood estimation does not feature an embedded variable selection procedure. 
  For\\
  GAMLSS models the standard AIC has been expanded to the \emph{generalized} AIC (GAIC) in \cite{rigby2005generalized} to be applied to multidimensional prediction functions. Variable selection via information criteria has several shortcomings, for example the inclusion of too many non-informative variables \cite{anderson2002avoiding}. 
\item[(3)] Whether to model predictors in a linear or nonlinear fashion is not trivial. 
  Unnecessary complexity increases the danger of overfitting as well as computation time. 
  Again, a generalized criterion like GAIC must be used to choose between linear and nonlinear terms.
\end{enumerate}


\subsection{Boosted GAMLSS}
\label{sec:boostedGAMLSS}

To deal with these issues, gradient boosting can be used to fit the model instead of the standard maximum likelihood algorithm.
Based on an approach proposed in \cite{schmid2010estimation} to fit zero-inflated count models, in \cite{mayr2012generalized} the author developed a general algorithm to fit multidimensional prediction functions with component-wise gradient boosting (see Algorithm~\ref{alg:comp_boost_mult}).

The basic idea is to cycle through the distribution parameters $\bm \theta$ in the fitting process. 
Partial derivatives with respect to each of the additive predictors are used as response vectors. 
In each iteration of the algorithm, the best-fitting base-learner is determined for \emph{each} distribution parameter while all other parameters stay fixed.
For a four parametric distribution, the update in boosting iteration $m + 1$ may be sketched as follows:
\begin{eqnarray*}
\frac{\partial}{\partial\eta_{\theta_1}}\rho(y, \hat\theta_1^{[m]}, \hat\theta_2^{[m]}, \hat\theta_3^{[m]}, \hat\theta_4^{[m]}) &\stackrel{\text{update}}{\longrightarrow}& \eta_{\theta_1}^{[m+1]} \\ 
\frac{\partial}{\partial\eta_{\theta_2}}\rho(y, \hat\theta_1^{[m+1]}, \hat\theta_2^{[m]}, \hat\theta_3^{[m]}, \hat\theta_4^{[m]}) &\stackrel{\text{update}}{\longrightarrow}& \eta_{\theta_2}^{[m+1]} \\
\frac{\partial}{\partial\eta_{\theta_3}}\rho(y, \hat\theta_1^{[m+1]}, \hat\theta_2^{[m + 1]}, \hat\theta_3^{[m]}, \hat\theta_4^{[m]}) &\stackrel{\text{update}}{\longrightarrow}& \eta_{\theta_3}^{[m+1]} \\ 
\frac{\partial}{\partial\eta_{\theta_4}}\rho(y, \hat\theta_1^{[m+1]}, \hat\theta_2^{[m + 1]}, \hat\theta_3^{[m + 1]}, \hat\theta_4^{[m]}) &\stackrel{\text{update}}{\longrightarrow}& \eta_{\theta_4}^{[m+1]}. \\ 
\end{eqnarray*}

Unfortunately, separate stopping values for each distribution parameter have to be specified, as the prediction functions will most likely require different levels of complexity and hence a different number of boosting iterations.
In case of multi-dimensional boosting the different $m_{\text{stop},k}$ values are not independent of each other, and have to be jointly optimized. 
The usually applied \emph{grid search} scales exponentially with the number of distribution parameters and can quickly become computationally demanding or even infeasible.  

\begin{algorithm}
\small
\caption{\enquote{Cyclical} component-wise gradient boosting in multiple dimensions \cite{mayr2012generalized}}
\label{alg:comp_boost_mult}
\vspace{3mm}
\textbf{Initialize}
\begin{enumerate}
\item[(1)] Initialize the additive predictors $\bm{\hat\eta}^{[0]}=(\hat\eta^{[0]}_{\theta_1}$, $\hat\eta^{[0]}_{\theta_2}$, $\hat\eta^{[0]}_{\theta_3}$, $\hat\eta^{[0]}_{\theta_4})$ with offset values.
\item[(2)] For each distribution parameter $\theta_k, k = 1,\dots, 4$, specify a set of base-learners, i.e., for parameter $\theta_k$ define $h_{k1}(\bm x^{(i)}), \dots, h_{kJ_k}(\bm x^{(i)})$ where $J_k$ is the cardinality of the set of base-learners specified for $\theta_k$.
\end{enumerate}
\textbf{Boosting in multiple dimensions}\\
For $m=1$ to $\max(m_{\text{stop},1},...,m_{\text{stop},4})$:
\begin{enumerate}
\item[(3)] For $k=1$ to $4$: 
\begin{itemize}
\item[(a)] \textbf{If} $m > m_{\text{stop},k}$ set $\hat{\eta}^{[m]}_{\theta_k} := \hat{\eta}^{[m-1]}_{\theta_k}$ and skip this iteration.\\
\textbf{Else} compute negative partial derivative $- \frac{\partial}{\partial\eta_{\theta_k}}\rho(y, \bm \eta)$ an plug in the current estimates $\bm{\hat\eta}^{[m-1]}(\cdot)$:
\[
\bm u_k = \left(\frac{\partial}{\partial\eta_{\theta_k}}\rho(y, \bm \eta)\Bigr|_{\bm\eta=\bm{\hat\eta}^{[m-1]}(\bm x^{(i)}), y=y^{(i)}}\right)_{i=1,\dots, n}
\]
\item[(b)] \textbf{Fit} each of the base-learners $\bm u_k$ contained in the set of base-learners specified for the distribution parameter $\theta_k$ in step (2) to the negative gradient vector.
\item[(c)] \textbf{Select} the component $j^*$ that best fits the negative partial-derivative vector according to the residual sum of squares, i.e., select the base-learner $h_{kj^*}$ defined by
\[
j^* = \argmin_{j \in 1,...,J_k}\sum^n_{i=1}(u_k^{(i)}-\hat{h}_{kj}(\bm x^{(i)}))^2.
\]
\item[(d)] \textbf{Update} the additive predictor $\eta_{\theta_k}$
\begin{equation*}
\hat\eta^{[m]}_{\theta_k} = \hat\eta^{[m-1]}_{\theta_k} + \stlen\cdot\hat{h}_{kj^*}(\bm x),
\end{equation*}
where $\stlen$ is the step-length (typically $\stlen = 0.1$), and update the current estimates for step 4(a): 
\begin{equation*}
\hat\eta^{[m - 1]}_{\theta_k} = \hat\eta^{[m]}_{\theta_k}.
\end{equation*}
\end{itemize}
\end{enumerate}
\end{algorithm}

\subsection{Stability selection}
\label{sec:stabsel}

Selecting an optimal subset of explanatory variables is a crucial step in almost every supervised data analysis problem. Especially in situations with a large number of covariates it is often almost impossible to get meaningful results without \emph{automatic}, or at least \emph{semi-automatic}, selection of the most relevant predictors. Selection of covariate subsets based on modified $R^2$ criteria (e.g., the $AIC$) can be unstable, see for example \cite{flack1987frequency}, and tend to select too many covariates (see, e.g., \cite{mayr2012importance}). 

Component-wise boosting algorithms are one solution to select predictors in high dimensions and/or $p > n$ problems. 
As only the best fitting base-learner is selected to update the model in each boosting step, as discussed above, variable selection can be obtained by stopping the algorithm early enough. 
Usually this is done via cross-validation methods, selecting the stopping iteration that optimizes the empirical risk on test data (\textit{predictive} risk). 
Hence, boosting with \emph{early stopping} via cross-validation offers a way to perform variable selection while fitting the model. 
Nonetheless,\\
boosted models stopped early via cross-validation still have a tendency to include too many noise variables, particularly in rather low-dimensional settings with few possible predictors and many observations ($n > p$) \cite{buhlmann2014discussion}.

\subsubsection{Stability selection for boosted GAM models}

To circumvent the problems mentioned above, the \emph{stability selection} approach was developed \cite{Meinshausen:Stab:2010,shah2013variable}. This generic algorithm can be applied to boosting and all other variable selection methods. The main idea of \emph{stability selection} is to run the selection algorithm on multiple subsamples of the original data. Highly relevant base-learners should be selected in (almost) all subsamples. 

Stability selection in combination with boosting was investigated in \cite{hofner2014controlling} and is outlined in Algorithm~\ref{alg:stabsel}. In the first step, $B$ random subsets of size $\lfloor n/2\rfloor$ are drawn and a boosting model is fitted to each one. The model fit is interrupted as soon as $q$ different base-learners have entered the model. For each base-learner the selection frequency $\hat{\pi}_j$ is the fraction of subsets in which the base-learner $j$ was selected \eqref{eq:sel_freq}. An effect is included in the model if the selection frequency exceeds the user-specified threshold $\pi_\text{thr}$ \eqref{eq:thr}.

\begin{algorithm}
\caption{Stability selection for model-based boosting}
\label{alg:stabsel}
\begin{enumerate}
\item For $b = 1,\dots, B$:
\begin{enumerate}
\item Draw a subset of size $\lfloor n/2\rfloor$ from the data
\item Fit a boosting model until the number of selected base-learners is equal to $q$ or the number of iterations reaches a pre-specified number ($m_\text{stop}$).\\
\end{enumerate}
\item Compute the relative selection frequencies per base-learner:
\begin{equation}
\label{eq:sel_freq}
\hat\pi_j := \frac{1}{B}\sum^B_{b=1}\mathds{I}_{\{j\in\hat S_{b}\}},
\end{equation}
where $\hat S_{b}$ denotes the set of selected base-learners in iteration $b$.
\item Select base-learners with a selection frequency of at least $\pi_\text{thr}$, which yields a set of stable covariates
\begin{equation}
\label{eq:thr}
\hat S_\text{stable} := \{j:\hat\pi_j\ge\pi_\text{thr}\}.
\end{equation}
\end{enumerate}
\end{algorithm}

This approach leads to upper bounds for the \emph{per-family error-rate} (PFER) $\mathds{E}(V)$, where  $V$ is the number of non-informative base-learners wrongly included in the model (i.e., false positives) \cite{Meinshausen:Stab:2010}:
\begin{equation}
\label{eq:error_bound}
\mathds{E}(V) \le \frac{q^2}{(2\pi_\text{thr}-1)p}.
\end{equation}
Under certain assumptions, refined, less conservative error bounds can be derived \cite{shah2013variable}.

One of the main difficulties of stability selection in practice is the choice of the parameters $q$, $\pi_\text{thr}$ and PFER. Even though only two of the three parameters need to be specified (the last one can then be derived under equality in \eqref{eq:error_bound}) their choice is not trivial and not always intuitive for the user. 

Meinshausen and Bühlmann \cite{Meinshausen:Stab:2010} state that the threshold should be $\pi_\text{thr} \in (0.6, 0.9)$ and has little influence on the result. The number of base-learners $q$ has to be sufficiently large, i.e., \@ $q$ should be at least as big as the number of informative variables in the data (or better to say the number of corresponding base-learners). This is obviously a problem in practical applications, in which the number of informative variables (or base-learners) is usually unknown. One nice property is that if $q$ is fixed, $\pi_\text{thr}$ and the PFER can be varied without the need to refit the model. A general advice would thus be to choose $q$ relatively large or to make sure that $q$ is large enough for a given combination of $\pi_\text{thr}$ and PFER. Simulation studies like \cite{hofner2014controlling,mayr2016stabsel} have shown that the PFER is quite conservative and the true number of false positives will most likely be much smaller than the specified value. 

In practical applications two different approaches to select the parameters are typically used. Both assume that the number of covariates to include, $q$, is chosen intuitively by the user: The first idea is to look at the calculated inclusion frequencies $\hat\pi_j$ and look for a \emph{breakpoint} in the decreasing order of the values. The threshold can be then chosen so that all covariates with inclusion frequencies larger than the breakpoint are included and the resulting PFER is only used as an additional information. The second possibility is to fix the PFER as a conservative upper bound for the expected number of false positives base-learners. Hofner \emph{et al.} \cite{hofner2014controlling} provide some rationales for the selection of the PFER by relating it to common error types, the per-comparison error (i.e., the type I error without multiplicity correction) and the family-wise error rate (i.e., with conservative multiplicity correction).

\subsubsection{Stability selection for boosted GAMLSS models}
\label{sec:stabs_for_gamboostLSS}

The question of variable selection in (boosted) GAMLSS models is even more critical than in regular (GAM) models, as the question of including a base-learner implies not only if the base-learner should be used in the model at all, but also for which distribution parameter(s) it should be used. Essentially, the number of possible base-learners doubles in a distribution with two parameters, triples in one with three parameters and so on. This is particularly challenging in situations with a large amount of base-learners and in $p > n$ situations.

The method of fitting boosted GAMLSS models in a cyclical way leads to a severe problem when used in combination with stability selection. In each iteration of the algorithm \emph{all} distribution parameters will receive an additional base-learner as long as their $m_\text{stop}$ limit is not exceeded. This means that base-learners are added to the model that might have a rather small importance compared to base-learners for other distribution parameters. This becomes especially relevant if the number of informative base-learners varies substantially between distribution parameters. 

Regarding the maximum number of base-learners $q$ to be considered in the model, base-learners are counted separately for each distribution parameter, so a base-learner that is selected for the location \emph{and} scale parameter counts as two different base-learners. Arguably, one might circumvent this problem by choosing a higher value for $q$, but still less stable base-learners could be selected instead of stable ones for other distribution parameters. One aspect of the problem is that the possible model improvement between different distribution parameters is not considered. A careful selection of $m_\text{stop}$ per distribution parameter might resolve the problem, but the process would still be unstable because the margin of base-learner selection in later stages of the algorithm is quite small. Furthermore, this is not in line with the general approach of stability selection where the standard tuning parameters do not play an important role.

\subsection{Noncyclical fitting for boosted GAMLSS}
\label{sec:noncycfitting}



The main idea to solve the previously stated problems of the cyclical fitting approach is to update only one distribution parameter in each iteration, i.e, the distribution parameter with a base-learner that leads to the highest loss reduction over all distribution parameters and base-learners. 

Usually, base-learners are selected by comparing their residual sum of squares with respect to the negative gradient vector (\emph{inner loss}). This is done in Step (4c) of Algorithm~\ref{alg:comp_boost_mult} where the different base-learners are compared. However, the residual sum of squares cannot be used to compare the fit of base-learners over different distribution parameters, as the gradients are not comparable.

\paragraph{Inner loss} 
One solution is to compare the empirical risks (i.e., the negative log likelihood of the modeled distribution) after the update with the best-fitting base-learners that have been selected via the residual sum of squares for each distribution parameter:  
First, for each distribution parameter the best performing base-learner is selected via the residual sum of squares of the base-learner fit with respect to the gradient vector.  Then, the potential improvement in the empirical loss $\Delta\rho$ is compared for all selected base-learners (i.e., over all distribution parameters). 
Finally, only the best-fitting base-learner (w.r.t. the inner-loss) which leads to the highest improvement (w.r.t. the outer loss) is updated. 
The base-learner selection for each distribution parameter is still done with the \emph{inner loss} (i.e., the residual sum of squares) and this algorithm will be called analogously.

\paragraph{Outer loss} 
Choosing base-learners and parameters with respect to two different optimization criteria may not always lead to the best possible update. 
A better solution could be to use a criterion which can compare all base-learners for all distribution parameters. 
As stated, the inner loss cannot be used for such a comparison.
However, the empirical loss (i.e., the negative \\
log-likelihood of the modeled distribution) can be used to compare both, the base-learners within a distribution parameter and over the different distribution parameters.
Now, the negative gradients are used to estimate all base-learners $\hat h_{11}, \dots, \hat h_{1p_1}, \hat h_{21},\dots, \hat h_{4p_4}$. 
The improvement in the empirical risk is then calculated for each base-learner of every distribution parameter and only the 
the overall best-performing base-learner (w.r.t. the outer loss) is updated. Instead of the using the inner loss, the whole selection process is hence based on the \emph{outer loss} (empirical risk) and the method is named accordingly.

The noncyclical fitting algorithm is shown in Algorithm~\ref{alg:comp_boost_mult_comb}. The \emph{inner} and \emph{outer} variant solely differ in step (3c).


\begin{algorithm}
\small
\caption{\enquote{Noncyclical} component-wise gradient boosting in multiple dimensions }
\label{alg:comp_boost_mult_comb}
\textbf{Initialize}
\begin{itemize}
\item[(1)] Initialize the additive predictors $\bm{\hat\eta}^{[0]}=(\hat\eta^{[0]}_{\theta_1}$, $\hat\eta^{[0]}_{\theta_2}$, $\hat\eta^{[0]}_{\theta_3}$, $\hat\eta^{[0]}_{\theta_4})$ with offset values.
\item[(2)] For each distribution parameter $\theta_k, k = 1,\dots, 4,$ specify a set of base-learners, i.e., for parameter $\theta_k$ define $h_{k1}(\cdot), \dots, h_{kJ_k}(\cdot)$ where $J_k$ is the cardinality of the set of base-learners specified for $\theta_k$.
\end{itemize}

\textbf{Boosting in multiple dimensions}\\
For $m = 1$ to $m_\text{stop}$:
\begin{enumerate}
\item[(3)] For $k = 1$ to $4$:
\begin{enumerate}
\item[(a)] Compute negative partial derivatives $- \frac{\partial}{\partial\eta_{\theta_k}}\rho(y, \bm \eta)$ and plug in the current estimates $\bm{\hat\eta}^{[m-1]}(\cdot)$:
\[
\bm u_k = \left(\frac{\partial}{\partial\eta_{\theta_k}}\rho(y, \bm \eta)\Bigr|_{\bm\eta=\bm{\hat\eta}^{[m-1]}(\bm x^{(i)}), y=y^{(i)}}\right)_{i=1,\dots, n}
\]
\item[(b)] \textbf{Fit} each of the base-learners $\bm u_k$ contained in the set of base-learners specified for the distribution parameter $\theta_k$ in step (2) to the negative gradient vector.
\item[(c)] \textbf{Select} the best-fitting base-learner $h_{kj^*}$ either by
\begin{itemize}
\item[$\bullet$] the inner loss, i.e., the residual sum of squares of the base-learner fit w.r.t. $\bm u_k$:
\begin{equation*}
j^* = \argmin_{j \in 1,\dots,J_k}\sum^n_{i=1}(u_{k}^{(i)}-\hat{h}_{kj}(\bm x^{(i)}))^2
\end{equation*}
\item[$\bullet$] the outer loss, i.e., the negative log likelihood of the modelled distribution after the potential update:
\begin{equation*}
j^* = \argmin_{j \in 1,\dots,J_k} \sum^n_{i=1}\rho\left(y^{(i)}, \hat\eta^{[m-1]}_{\theta_k}(\bm x^{(i)}) + \stlen\cdot\hat{h}_{kj}(\bm x^{(i)})\right)
\end{equation*}
\end{itemize}
\item[(d)] Compute the possible improvement of this update regarding the outer loss
\begin{equation*}
\Delta\rho_k = \sum^n_{i=1}\rho\left(y^{(i)}, \hat\eta^{[m-1]}_{\theta_k}(\bm x^{(i)}) + \stlen\cdot\hat{h}_{kj^*}(\bm x^{(i)})\right)
\end{equation*}
\end{enumerate}
\item[(4)] \textbf{Update}, depending on the value of the loss reduction \\
$k^*=\argmin_{k \in 1,\dots,4}(\Delta\rho_k)$ only the overall best-fitting base-learner:
\begin{equation*}
\hat\eta^{[m]}_{\theta_{k^*}}=\hat\eta^{[m-1]}_{\theta_{k^*}} + \stlen\cdot\hat{h}_{k^*j^*}(\bm x)
\end{equation*}
\item[(5)] Set $\hat\eta^{[m]}_{\theta_{k}} := \hat\eta^{[m-1]}_{\theta_{k}}$ for all $k \ne k^*$.
\end{enumerate}
\end{algorithm}

A major advantage of both noncyclical variants compared to the cyclical fitting algorithm (Algorithm~\ref{alg:comp_boost_mult}) is that $m_\text{stop}$ is always scalar.
The updates of each distribution parameter estimate are adaptively chosen. The optimal partitioning (and sequence) of base-learners between different parameters is done automatically while fitting the model. Such a scalar optimization can be done very efficiently using standard cross-validation methods without the need for a multi-dimensional grid search. 

\section{Simulation study}
\label{sec:simulation}

In a first step, we carry out simulations to evaluate the performance of the new noncyclical fitting algorithms regarding convergence, convergence speed and runtime. In a second step, we analyze the variable selection properties if the new variant is combined with stability selection.

\subsection{Performance of the noncyclical algorithms}
The response $y_i$ is drawn from a normal distribution $N(\mu_i,\sigma_i)$, where $\mu_i$ and $\sigma_i$ depend on $4$ covariates each. The $x_i, i = 1, \ldots, 6,$ are drawn independently from a uniform distribution on $[-1, 1]$, i.e., $n = 500$ samples are drawn independently from $U(-1,1)$. Two covariates $x_3$ and $x_4$ are shared between both $\mu_i$ and $\sigma_i$, i.e., they are informative for both parameters, which means that there are $p_{\text{inf}}=6$ informative variables overall. The resulting predictors look like 
\begin{description}
\item[ ] $\mu_i = x_{1i} + 2x_{2i} + 0.5x_{3i} - x_{4i}$
\item[ ] $\log(\sigma_i) = 0.5x_{3i} + 0.25x_{4i} - 0.25x_{5i} - 0.5x_{6i}$.
\end{description}

\paragraph{Convergence} First, we compare the new noncyclical\\ 
boosting algorithms and the cyclical approach with the classical estimation method based on penalized maximum likelihood (as implemented in the R package\\
\textbf{gamlss}, \cite{stasinopoulos2008Instructions}). The results from $B = 100$ simulation runs are shown in Figure~\ref{fig:sim_conv}. All four methods converge to the correct solution.

\begin{knitrout}
\definecolor{shadecolor}{rgb}{0.969, 0.969, 0.969}\color{fgcolor}\begin{figure}

\includegraphics[width=\maxwidth]{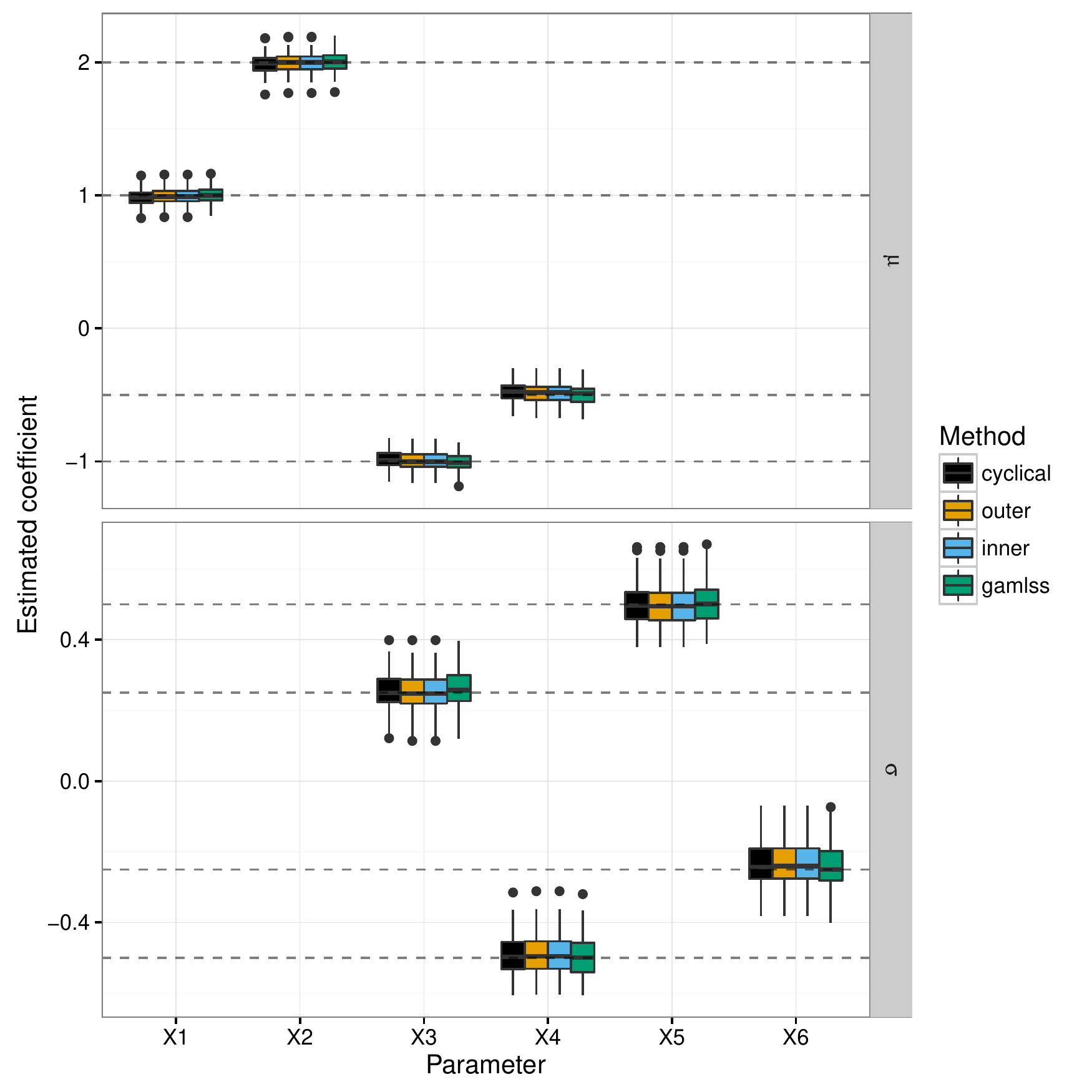} \hfill{}

\caption[Distribution of coefficient estimates from $B = 100$ simulation runs]{Distribution of coefficient estimates from $B = 100$ simulation runs. The dashed lines show the true parameters. All algorithms were fitted until convergence.}\label{fig:sim_conv}
\end{figure}

\end{knitrout}

\paragraph{Convergence speed} Second, we compare the convergence speed in terms of boosting iterations. Therefore, non-informative variables are added to the model. Four settings are considered with $p_{\text{n-inf}}$ $= 0,50,250$ and $500$ additional non-informative covariates independently sampled from a U(-1,1) distribution. With $n=500$ observations, both $p_{\text{n-inf}}=250$ and $p_{\text{n-inf}}=500$ are high-dimensional situations ($p > n$) as we have two distribution parameters. In Figure \ref{fig:conv_speed_norm} the mean risk over $100$ simulated data sets is plotted against the number of iterations. The $m_\text{stop}$ value of the cyclical variant shown in Figure \ref{fig:conv_speed_norm} is the sum of the number of updates on every distribution parameter. Outer and inner loss variants of the noncyclical algorithm have exactly the same risk profiles in all four settings. Compared to the cyclical algorithm, the convergence is faster in the first $500$ iterations. After more than $500$ iterations the risk reduction is the same for all three methods. The margin between  cyclical and both noncyclical algorithms decreases with a larger number of noise variables.  

\begin{knitrout}
\definecolor{shadecolor}{rgb}{0.969, 0.969, 0.969}\color{fgcolor}\begin{figure}
\includegraphics[width=\maxwidth]{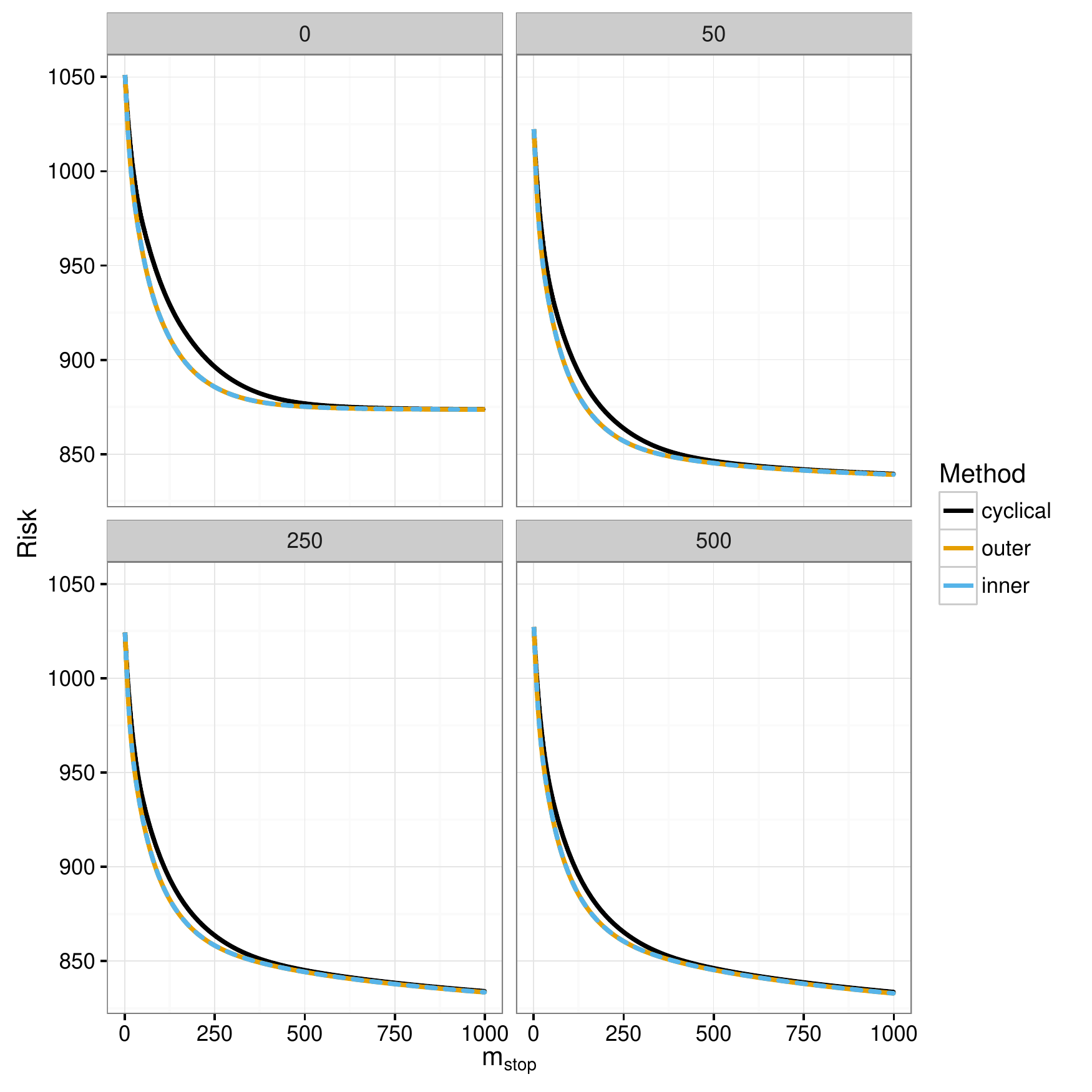} \caption[Convergence speed (regarding the number of boosting iterations $m$) with 6 informative and $p_{\text{n-inf}}= 0, 50, 250$ and $500$ additional noise variables]{Convergence speed (regarding the number of boosting iterations $m$) with 6 informative and $p_{\text{n-inf}}= 0, 50, 250$ and $500$ additional noise variables.}\label{fig:conv_speed_norm}
\end{figure}

\end{knitrout}

\paragraph{Runtime} The main computational effort of the algorithms is the base-learner selection, which is different for all three methods. 
The runtime is evaluated in context of cross-validation, which allows us to see how out-of-bag error and runtime behave in different settings. We consider two scenarios --- a two-dimensional ($d = 2$) and a three-dimensional ($d = 3$) distribution. The data are generated according to setting 1A and 3A of Section~\ref{sec:stabs_sim}. In each scenario we sample $n = 500$ observations, but do not add any additional noise variables. For optimization of the model, the out-of-bag prediction error is estimated via a $25$-fold bootstrap. A grid of length $10$ is created for the cyclical model, with an maximum $m_\text{stop}$ of $300$ for each distribution parameter. The grid is created with the \texttt{make.grid} function in \textbf{gamboostLSS} (refer to the package documentation for details on the arrangement of the grid points). To allow the same complexity for all variants, the noncyclical methods are allowed up to $m_\text{stop} = d \times 300$ iterations.

\begin{knitrout}
\definecolor{shadecolor}{rgb}{0.969, 0.969, 0.969}\color{fgcolor}\begin{figure}
\includegraphics[width=\maxwidth]{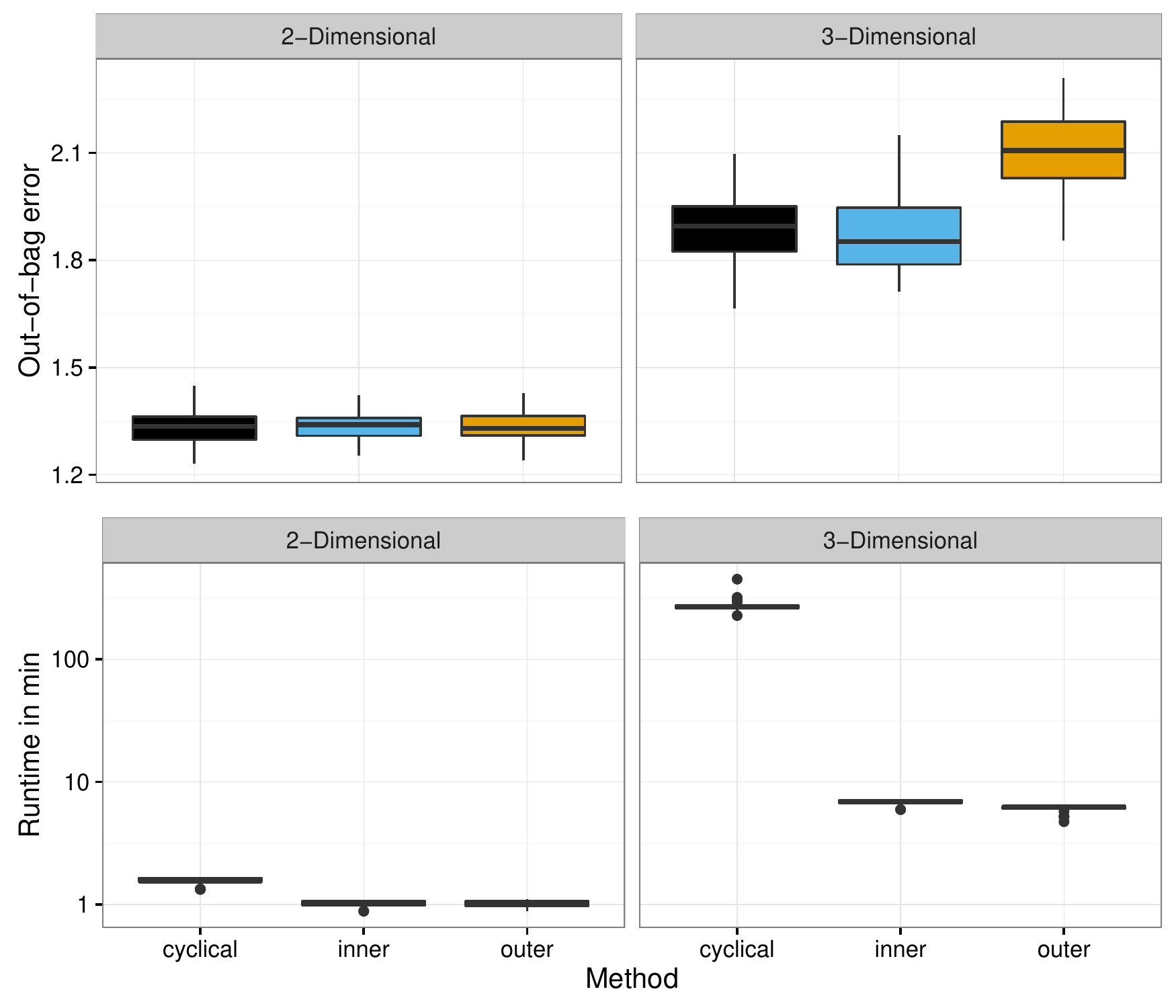} \caption[Out-of-bag error (top) and optimization time in minutes (logarithmic scale]{Out-of-bag error (top) and optimization time in minutes (logarithmic scale; bottom) for a two-dimensional (left) and three-dimensional distribution (right) based on $25$-fold bootstrap.}\label{fig:bench_optim}
\end{figure}

\end{knitrout}

The results of the benchmark can be seen in Figure~\ref{fig:bench_optim}. 
The out-of-bag error in the two-dimensional setting is similar for all three methods, but the average number of optimal iterations is considerably smaller for the noncyclical methods (\texttt{cyclical:}$360$ vs. \texttt{inner:}$306$,\\
\texttt{outer:}$308$). In the three-dimensional setting, the outer variant of the noncyclical fitting results in a higher error, whereas the inner variant results in a slightly better performance compared to the cyclical variant. In this setting the optimal number of iterations is similar for all three methods but near the edge of the searched grid. It is possible that the outer variant will result in a comparable out-of-bag error if the range of the grid is increased.


\subsection{Stability selection}
\label{sec:stabs_sim}
After having analyzed the properties of the new noncyclical boosting algorithms for fitting GAMLSS, the remaining question is how they perform when combined with stability selection. In the previous subsection no differences in the model fit (Figure~\ref{fig:sim_conv}) and convergence speed (Figure~\ref{fig:conv_speed_norm}) could be observed, but the optimization results in a three dimensions setting (Figure~\ref{fig:bench_optim}) was worse for the outer algorithm. Taking this into consideration we will only compare the inner and cyclical algorithm here. 

We consider three different distributions: (1) The \textit{normal} distribution with two parameters, mean $\mu_i$ and standard deviation $\sigma_i$.  (2) The \textit{negative binomial} distribution with two parameters, mean $\mu_i$ and dispersion $\sigma_i$. (3) The \textit{zero-inflated negative binomial} (ZINB) distribution with three parameters, $\mu_i$ and $\sigma_i$ identical to the negative binomial distribution, and probability for zero-inflation $\nu_i$.

Furthermore, two different partitions of six informative covariates shared between the distribution parameters are evaluated:
\begin{enumerate}
\item[{(A)}] \textit{Balanced case}: For normal and negative binomial distribution, both $\mu_i$ and $\sigma_i$ depended on four informative covariates, where two are shared. In case of the ZINB distribution, each parameter depends on three informative covariates, each sharing one with the other two parameters. 
\item[{(B)}] \textit{Unbalanced case}: For normal and negative binomial distribution, $\mu_i$ depends on five informative covariates, while $\sigma_i$ only on one. No informative variables are shared between the two parameters. For the ZINB distribution, $\mu_i$ depends on five informative variables, $\sigma_i$ on two, and $\nu_i$ on one. One variable is shared across all three parameters.  
\end{enumerate}

To summarize these different scenarios for a total of six informative variables, $x_1,\dots,x_6$:
\begin{description}
\item[(1A, 2A)]
\item[ ] $\mu_i = \beta_{1\mu}x_{1i} +\beta_{2\mu}x_{2i} + \beta_{3\mu}x_{3i} + \beta_{4\mu}x_{4i}$
\item[ ] $\log(\sigma_i) = \beta_{3\sigma}x_{3i} +\beta_{4\sigma}x_{4i} + \beta_{5\sigma}x_{5i} + \beta_{6\sigma}x_{6i}$\\

\item[(1B, 2B)]
\item[ ] $\log(\mu_i) = \beta_{1\mu}x_{1i} +\beta_{2\mu}x_{2i} + \beta_{3\mu}x_{3i} + \beta_{4\mu}x_{4i} + \beta_{5\mu}x_{5i}$
\item[ ] $\log(\sigma_i) = \beta_{6\sigma}x_{6i}$\\

\item[(3A)]
\item[ ] $\log(\mu_i) = \beta_{1\mu}x_{1i} +\beta_{2\mu}x_{2i} + \beta_{3\mu}x_{3i}$
\item[ ] $\log(\sigma_i) = \beta_{3\sigma}x_{3i} +\beta_{4\sigma}x_{4i} + \beta_{5\sigma}x_{5i}$
\item[ ] $\text{logit}(\nu_i) = \beta_{1\nu}x_{1i} + \beta_{5\nu}x_{5i} + \beta_{6\nu}x_{6i}$\\

\item[(3B)]
\item[ ] $\log(\mu_i) = \beta_{1\mu}x_{1i} +\beta_{2\mu}x_{2i} + \beta_{3\mu}x_{3i} + \beta_{4\mu}x_{4i} + \beta_{5\mu}x_{5i}$
\item[ ] $\log(\sigma_i) = \beta_{5\sigma}x_{5i} + \beta_{6\sigma}x_{6i}$
\item[ ] $\text{logit}(\nu_i) = \beta_{6\nu}x_{6i}$
\end{description}

To evaluate the performance of stability selection, two criteria have to be considered. First, the \emph{true positive rate}, or the number of \emph{true positives} (TP, number of correctly identified informative variable). Secondly, the \emph{false positive rate}, or the number of \emph{false positives} (FP, number of non-informative variable that were selected as stable predictors). 

Considering stability selection, the most obvious control parameter to influence false and true positive rates is the threshold $\pi_\text{thr}$. To evaluate the algorithms depending on the settings of stability selection, we consider several values for the number of variables to be included in the model $q \in \{8,15,25,50\}$ and the threshold $\pi_\text{thr}$ (varying between $0.55$ and $0.99$ in steps of $0.01$). A third factor is the number of (noise) variables in the model: we consider $p = 50, 250$ or $500$ covariates (including the six informative ones). It should be noted that the actual number of possible base-learners is $p$ times the number of distribution parameters, as each covariate can be included in one or more additive predictors. To visualize the simulation results, the progress of true and false positives is plotted against the threshold $\pi_\text{thr}$ for different values of $p$ and $q$, where true and false positives are aggregated over all distribution parameters. Separate figures for each distribution parameter can be found in the web supplement. The setting $p=50, q=50$ is an edge case that would work for some assumptions about the distribution of selection probabilities \cite{shah2013variable}. Since the practical application of this scenario is doubtful, we will not further examine it here.


\begin{knitrout}
\definecolor{shadecolor}{rgb}{0.969, 0.969, 0.969}\color{fgcolor}\begin{figure}
\includegraphics[width=\maxwidth]{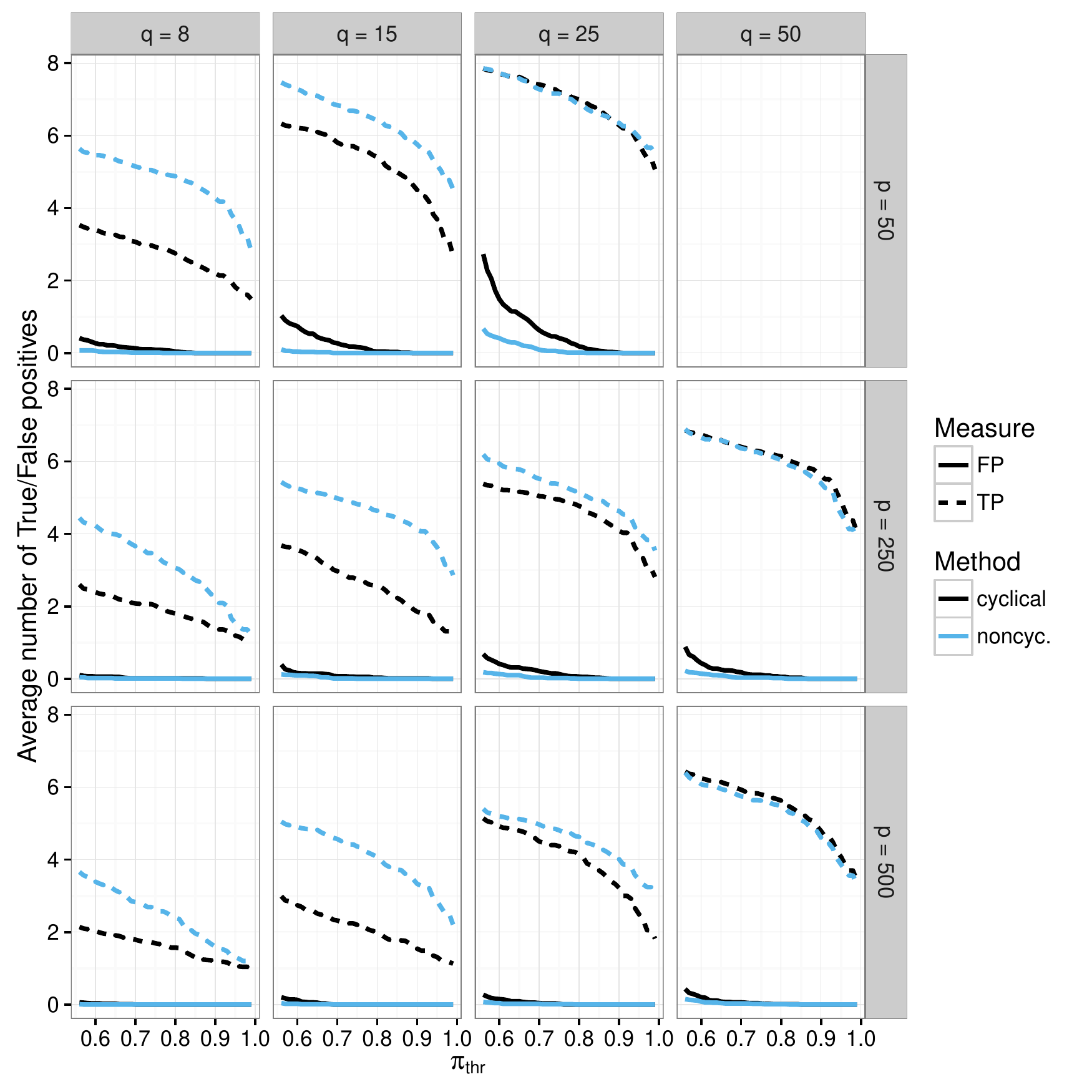} \caption[Balanced case with normal distribution (Scenario 1A)]{Balanced case with normal distribution (Scenario 1A).}\label{fig:stabs_norm_bal}
\end{figure}

\end{knitrout}

\begin{knitrout}
\definecolor{shadecolor}{rgb}{0.969, 0.969, 0.969}\color{fgcolor}\begin{figure}
\includegraphics[width=\maxwidth]{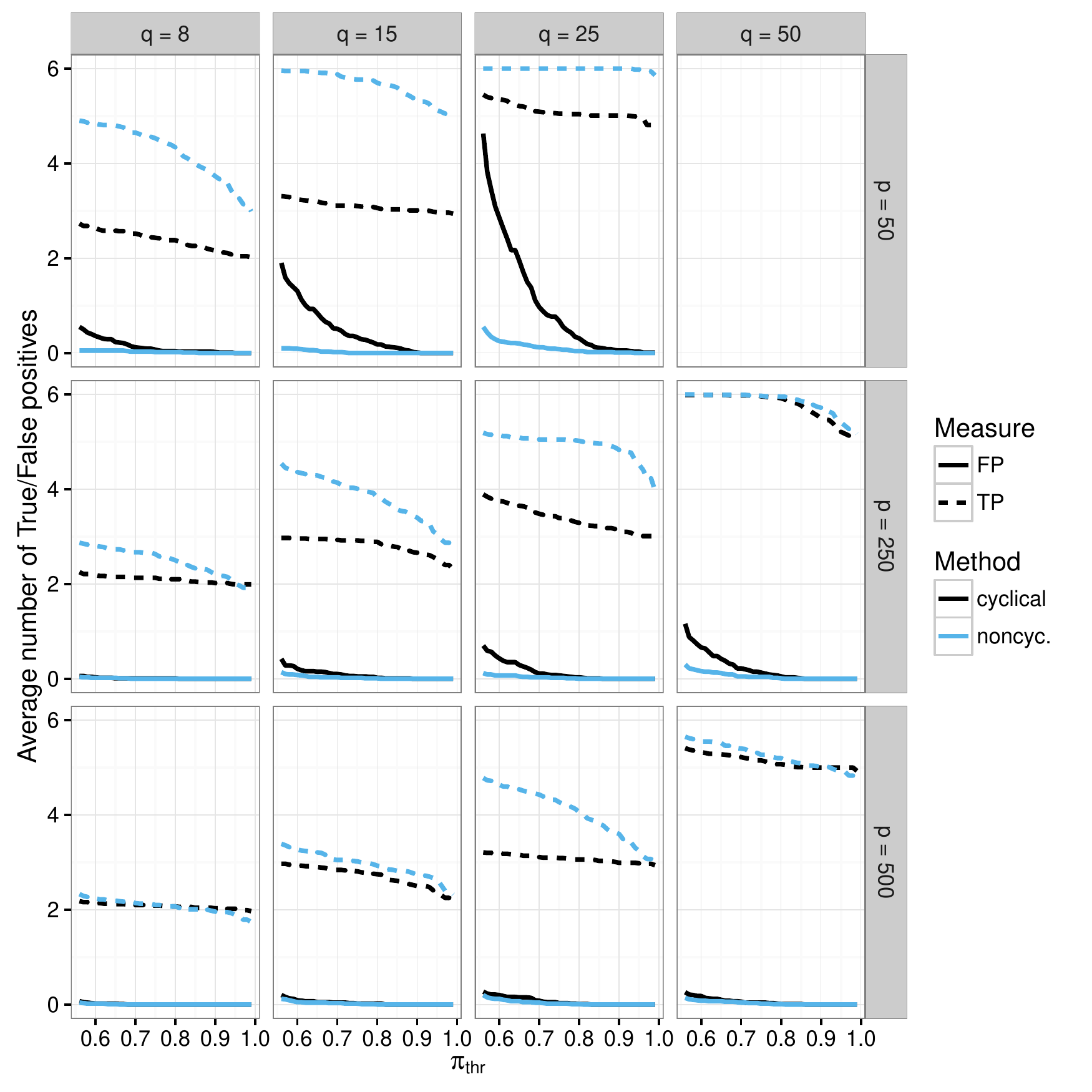} \caption[Unbalanced case with normal distribution (Scenario 1B)]{Unbalanced case with normal distribution (Scenario 1B).}\label{fig:stabs_norm_unbal}
\end{figure}

\end{knitrout}

\subsubsection{Results}

It can be observed that with increasing threshold $\pi_\text{thr}$, the number of true positives as well as the number of false positives declines in all six scenarios (see Figures~\ref{fig:stabs_norm_bal} to~\ref{fig:stabs_zinb_unbal}) and for every combination of $p$ and $q$. This is a natural consequence as the threshold is increased, the less variables are selected. Furthermore, the PFER, which is to be controlled by stability selection, decreases with increasing threshold $\pi_\text{thr}$ (see Eq.~\ref{eq:error_bound}).

\paragraph{Results for the normal distribution}

In the balanced case (Figure~\ref{fig:stabs_norm_bal}) a higher number of true positives for the noncyclical algorithm can be observed compared to the cyclical algorithm for most simulation settings. Particularly for smaller $q$ values ($q\in \{8,15\}$) the true positive rate was always higher compared to the cyclical variant. For higher $q$ values the margin decreases and for the highest settings both methods have approximately the same progression over $\pi_\text{thr}$, with slightly better results for the cyclical algorithm. Overall, the number of true positives increases with a higher value of $q$. Hofner \emph{et al.}~\cite{hofner2014controlling} found similar results for boosting with one dimensional prediction functions, but also showed that the true positive rate decreases again after a certain value of $q$. This could not be verified for the multidimensional case.

The false positive rate is extremely low for both methods, especially in the high-dimensional settings. The noncyclical fitting method has a constantly smaller or identical false positive rate and the difference reduces for higher $\pi_\text{thr}$, as expected. For all settings the false positive rate reaches zero for a threshold higher than $0.9$. The setting with the highest false positive rate is $p=50$ and $q=25$, a low dimensional case with a relatively high threshold. This is also the only setting where on average all $8$ informative variables are found (for a threshold of $0.55$).

In the unbalanced case (Figure~\ref{fig:stabs_norm_unbal}) the results are similar. The number of false positives for the noncyclical variant is lower compared to the cyclical approach in almost all settings. The main difference between the balanced and the unbalanced case is that the number of true positives for the $p=50, q=25$ setting is almost identical in the former case whereas in the latter case the noncyclical variant is dominating the cyclical algorithm. On the other hand, in the high-dimensional case with a small $q$ ($p=500, q = 8)$ both fitting methods have about the same true positive rate for all possible threshold values.

In summary, it can be seen that the novel noncyclical algorithm is generally better, but at least comparable, to the cyclical method in identifying informative variables. Furthermore, the false positive rate is less or identical to the cyclical method. For some scenarios in which the scale parameter $\sigma_i$ is higher compared to the location parameter $\mu_i$, the cyclical variant achieves slightly better results than the noncyclical variant regarding true positives at high $p$ and $q$ values.

\paragraph{Results for the negative binomial distribution}

\begin{knitrout}
\definecolor{shadecolor}{rgb}{0.969, 0.969, 0.969}\color{fgcolor}\begin{figure}
\includegraphics[width=\maxwidth]{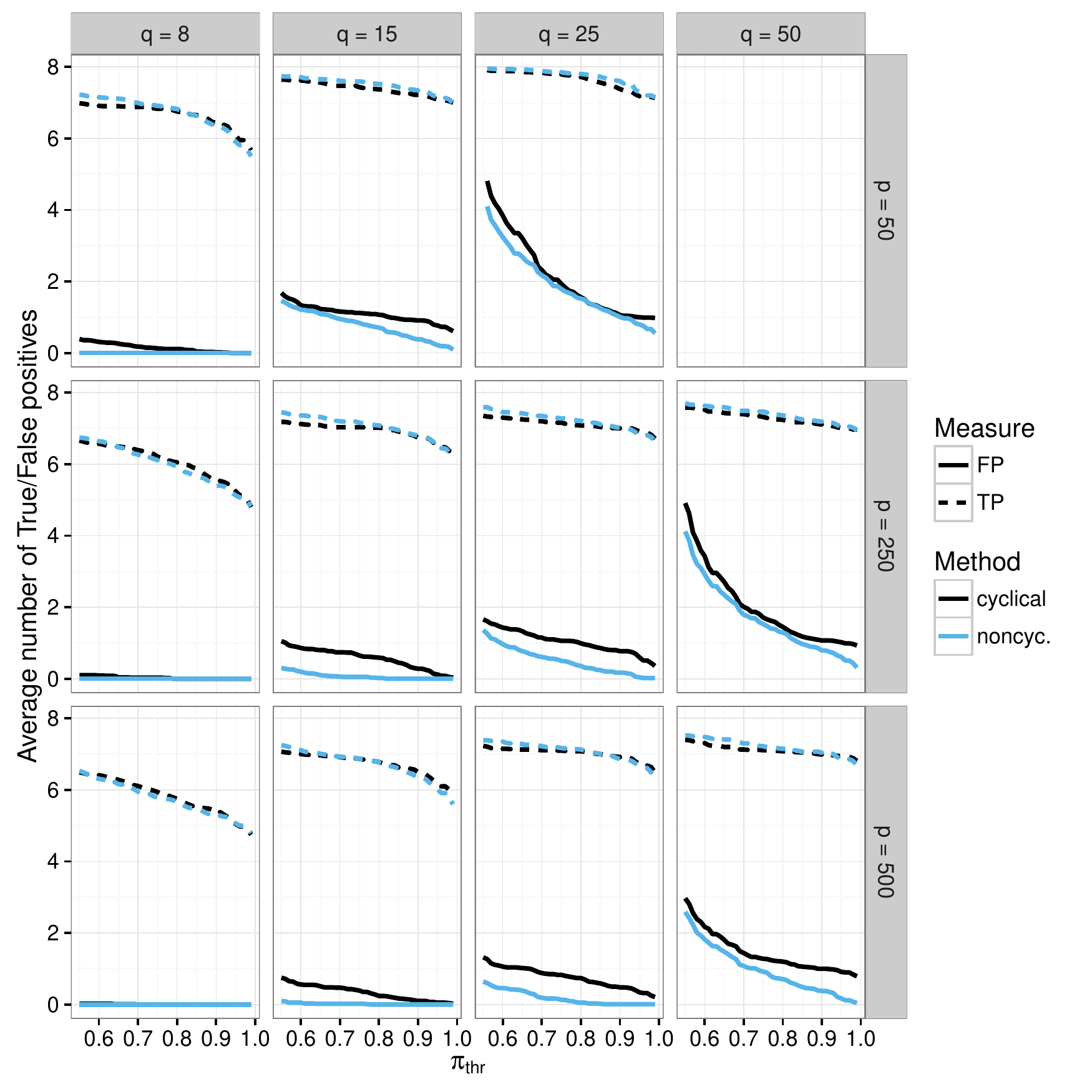} \caption[Balanced case with negative binomial distribution (Scenario 2A)]{Balanced case with negative binomial distribution (Scenario 2A).}\label{fig:stabs_nb_bal}
\end{figure}

\end{knitrout}

\begin{knitrout}
\definecolor{shadecolor}{rgb}{0.969, 0.969, 0.969}\color{fgcolor}\begin{figure}
\includegraphics[width=\maxwidth]{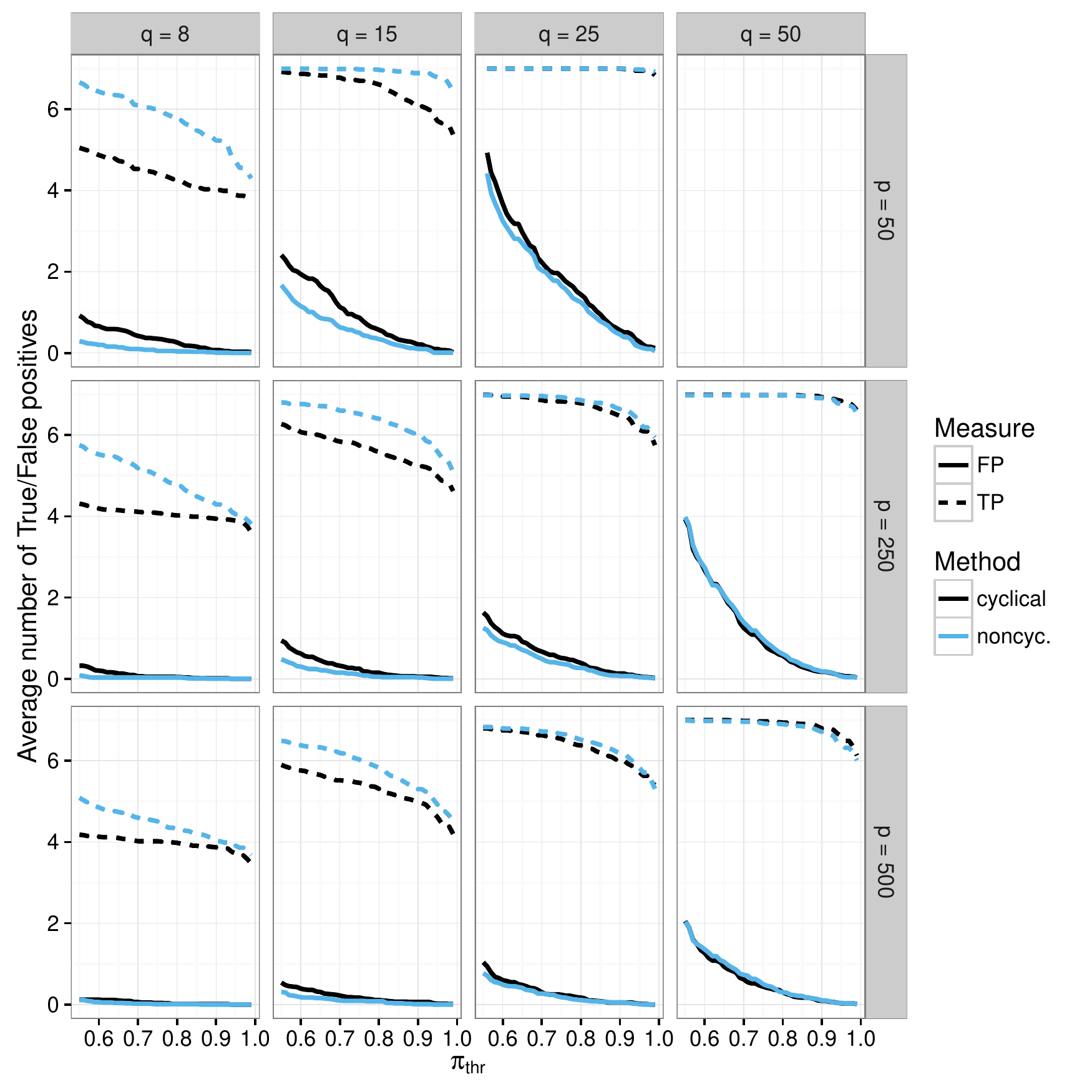} \caption[Unbalanced case with negative binomial distribution (Scenario 2B)]{Unbalanced case with negative binomial distribution (Scenario 2B).}\label{fig:stabs_nb_unbal}
\end{figure}

\end{knitrout}

In the balanced case of the negative binomial distribution (Figure~\ref{fig:stabs_nb_bal}), the number of true positives is almost identical for the cyclical and noncyclical algorithm in all settings, while the number of true positives is generally quite high. It varies between  $6$ and $8$ in almost all settings, except for the cases with a very small value of $q$ ($= 8$) where it is slightly lower. This is consistent with the results for stability selection with one dimensional boosting \cite{hofner2014controlling,mayr2016stabsel}. The number of false positives in the noncyclical variants is smaller or identical to the cyclical variant in all tested settings.

In the unbalanced case the true positive rate of the noncyclical variant is higher compared to the cyclical variant, whereas the difference reduces for larger values of $q$. The results are consistent with the normal distribution setting but with smaller differences between both methods.


\paragraph{Results for ZINB distribution}

The third considered distribution in our simulation setting is the ZINB distribution, which features three parameters to fit.

In Figure~\ref{fig:stabs_zinb_bal}, the results for the balanced case (scenario 3A), are visualized. The tendency of a larger number true positives in the noncyclical variant, which could be observed for both two-parametric distributions, is not present here. For all settings, except for high dimensional settings with a low $q$ (i.e., $p=250,500$ and $q=50$), the cyclical variant has a higher number of true positives. Additionally, the number of false positives is constantly higher for the noncyclical variant. For the unbalanced setting (Figure~\ref{fig:stabs_zinb_unbal}) the results are similar in true positives and negatives between both methods. 

\begin{knitrout}
\definecolor{shadecolor}{rgb}{0.969, 0.969, 0.969}\color{fgcolor}\begin{figure}
\includegraphics[width=\maxwidth]{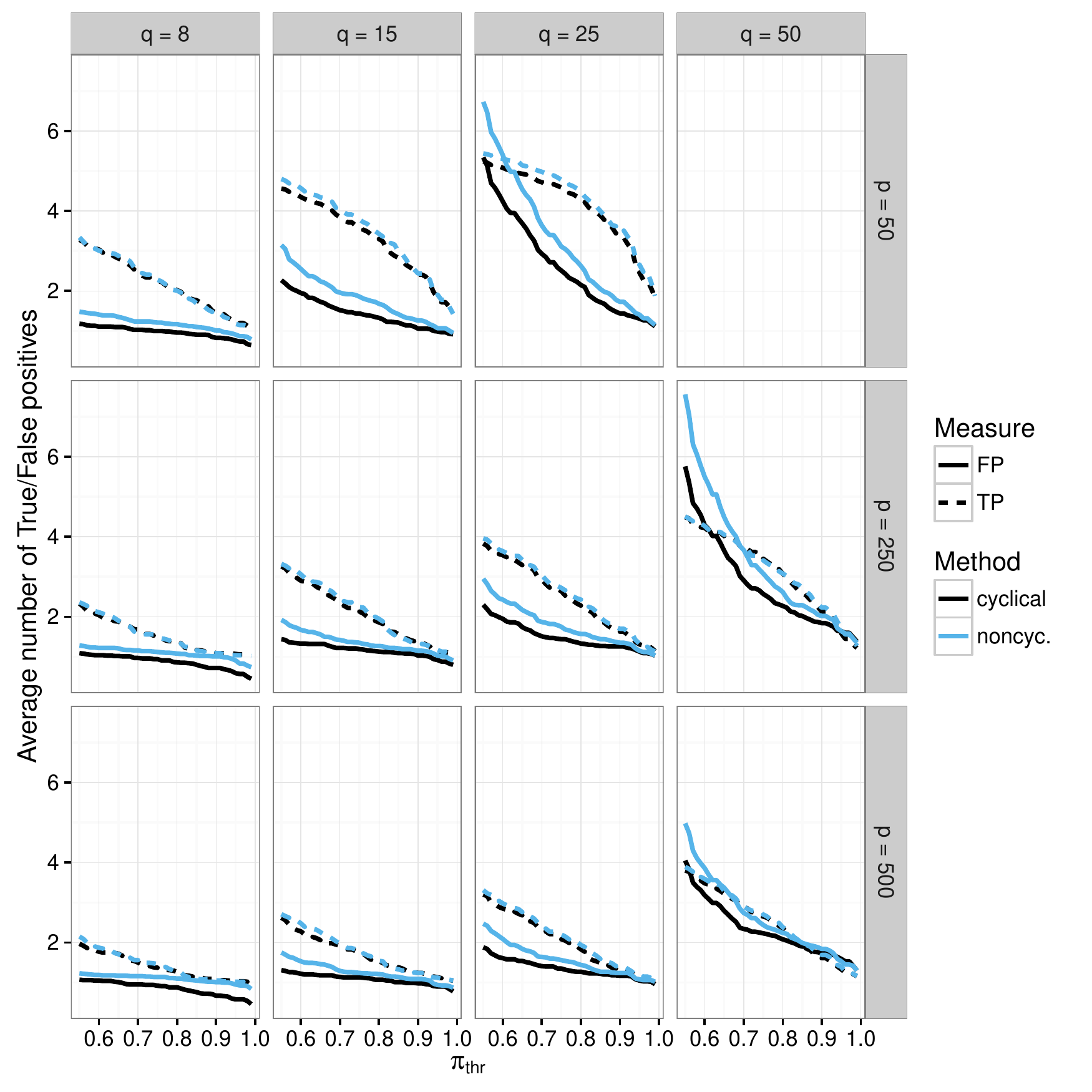} \caption[Balanced case with zero-inflated negative binomial distribution (Scenario 3A)]{Balanced case with zero-inflated negative binomial distribution (Scenario 3A).}\label{fig:stabs_zinb_bal}
\end{figure}

\end{knitrout}

\begin{knitrout}
\definecolor{shadecolor}{rgb}{0.969, 0.969, 0.969}\color{fgcolor}\begin{figure}
\includegraphics[width=\maxwidth]{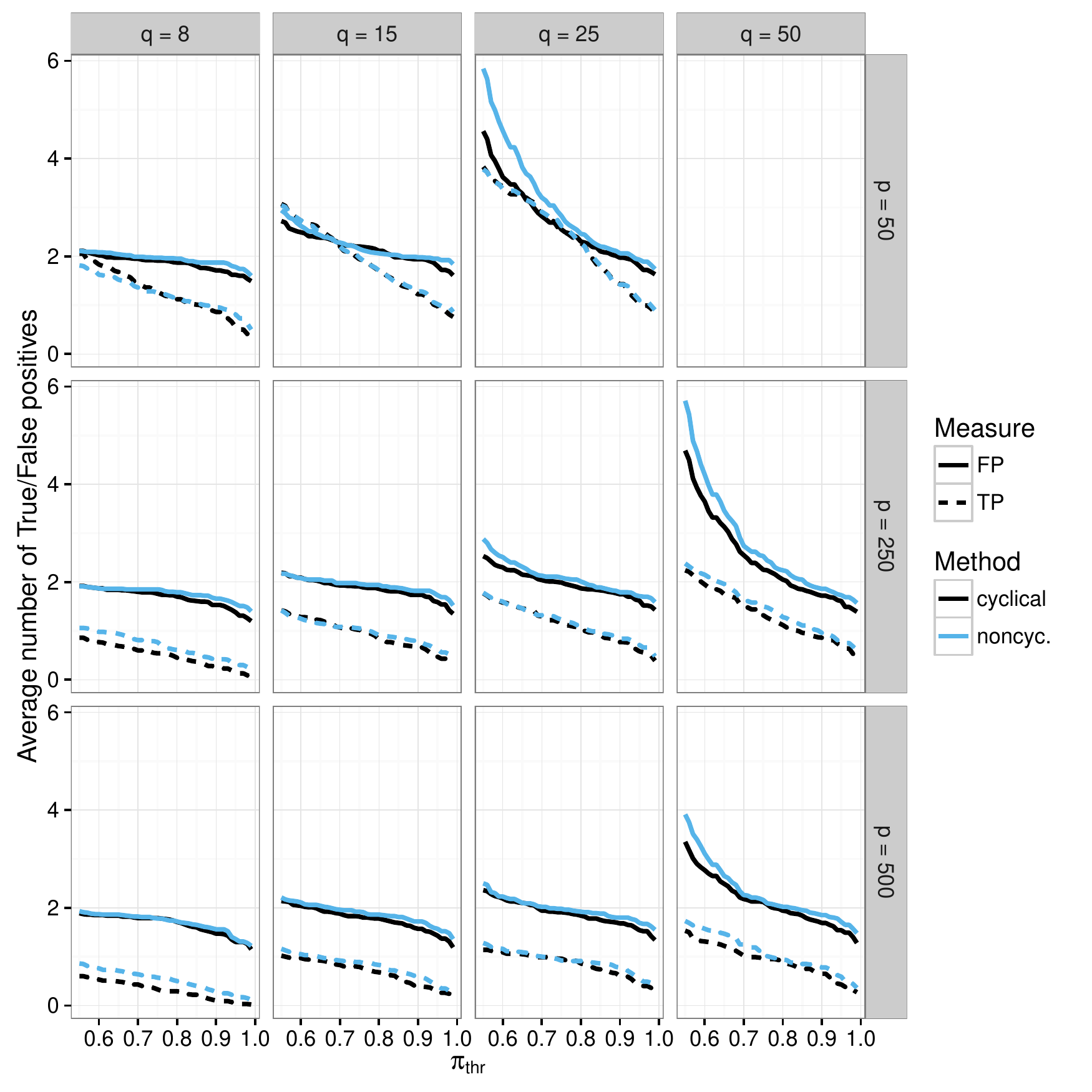} \caption[Unbalanced case with zero-inflated negative binomial distribution (Scenario 3B)]{Unbalanced case with zero-inflated negative binomial distribution (Scenario 3B).}\label{fig:stabs_zinb_unbal}
\end{figure}

\end{knitrout}

The number of true positives is overall considerably smaller compared to all other simulation settings. Particularly in the high dimensional cases ($p=250, 500$), not even half of the informative covariates are found. In settings with smaller $q$ the number of true positives is lower than two. Both algorithms obtain approximately the same number of true positives for all settings. In cases with a very low or a very high number $q$, (i.e., $q = 8$ or $50$), the noncyclical algorithm is slightly better. The number of false positives is very high, especially compared with the number of true positives and particularly for the unbalanced case. For a lot of settings, more than half of the included variables are non-informative. The number of false positives is higher for the noncyclical case. The difference are especially present in settings with a high $q$ and a low $\pi_\text{thr}$, those settings which also have the highest numbers of true positives.

Altogether, the trend from the simulated two-parameter distributions is not present in the three parametric setting. The cyclical algorithm overall is not worse or even better with regard to both true and false positives for almost all tested scenarios. 


\section{Modelling sea duck abundance}
\label{sec:application}

\begin{figure}
\centering
\includegraphics{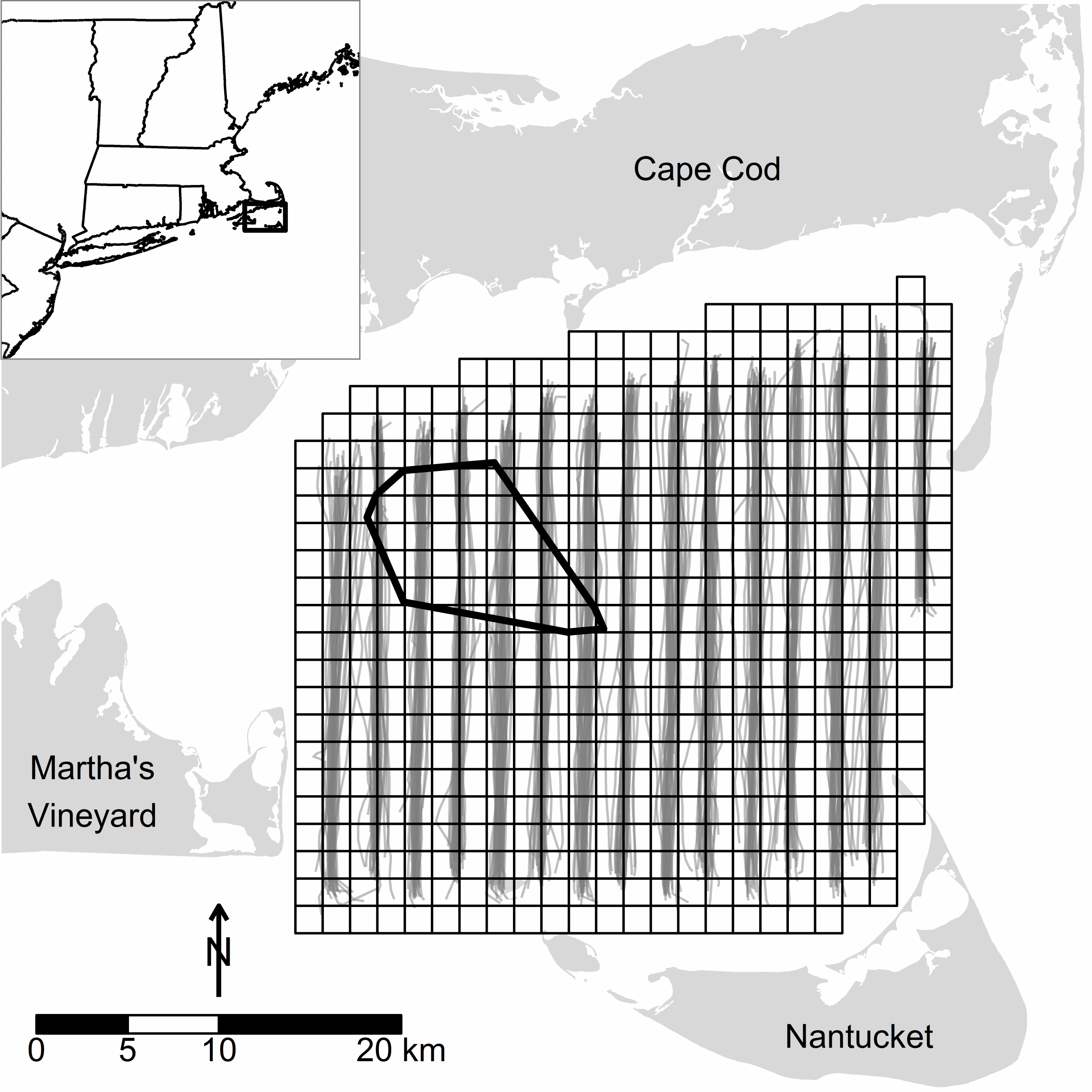}
\caption{Nantucket Sound -- Research area of the seabird study by Smith \emph{et al.}~\cite{smith2016seabirds}. Squares are the discretized segments in which bird abundance was studied. Gray lines indicate all aerial transects flown over the course of the study. The black polygon indicates the location of permitted wind energy
development on Horseshoe Shoal.}
\label{fig:birds}
\end{figure}

A recent analysis by Smith \emph{et al.}~\cite{smith2016seabirds} investigated the abundance of wintering sea ducks in Nantucket Sound, Massachusetts, USA. Spatio-temporal abundance data for common eider (among other species) was collected between 2003 and 2005 by counting seaducks on multiple aerial strip transects from a small plane. For the subsequent analysis, the research area was split in\\
$2.25km^2$ segments (see Figure~\ref{fig:birds}). Researchers were interested in variables that explained and predicted the distribution of the common eider in the examined area.

As the data were zero-inflated ($75 \%$ of the segments contained no birds) and highly skewed (a small
number of segments contained up to $30000$ birds), a hurdle model \cite{mullahy1986specification} was used for estimation. Therefore, the model was split into an occupancy model (zero part) and an abundance model (count part). The occupancy model estimated if a segment was populated at all and was fitted by boosting a generalized additive model (GAM) with binomial loss, i.e., an additive logistic regression model. In the second step, the number of birds in populated segments was estimated with a boosted GAMLSS model. Because of the skewed and long-tailed data, the (zero-truncated) \emph{negative binomial} distribution was chosen for the abundance model (compare \cite{mullahy1986specification}).

We reproduce the common eider model reported by Smith \emph{et al.}~\cite{smith2016seabirds} but apply the novel noncyclical algorithm; Smith \emph{et al.} used the cyclic algorithm to fit the GAMLSS model.  As discussed in Section~\ref{sec:stabs_sim} we apply the noncyclical algorithm with inner loss. In short, both distribution parameters, mean \emph{and} overdispersion of the abundance model, and the probability of bird sightings in the occupancy model were regressed on a large number of biophysical covariates, spatial and spatio-temporal effects, and some pre-defined interactions. A complete list of the considered effects can be found in the web supplement. To allow model selection (i.e., the selection between modelling alternatives), the covariate effects were split in linear and nonlinear base-learners \cite{Hothorn:2011:EcoMonographs,Hofner:unbiased:2011}. 
The step-length was set to $\stlen = 0.3$ and the optimal number of boosting iterations $m_\text{stop}$ was found via 25-fold subsampling with sample size $n/2$ \cite{mayr2012importance}. Additionally, we used stability selection to obtain sparser models. The numbers of variables to be included per boosting run was set to $q=35$ and the per-family error-rate was set to $6$. With unimodality assumption this resulted in a threshold of $\pi_\text{thr} = 0.9$. These settings were chosen identically to the original choices in \cite{smith2016seabirds}.

\subsection{Results}

\begin{figure}
\centering
\begin{subfigure}{.5\textwidth}
\centering
\includegraphics[width=.8\linewidth]{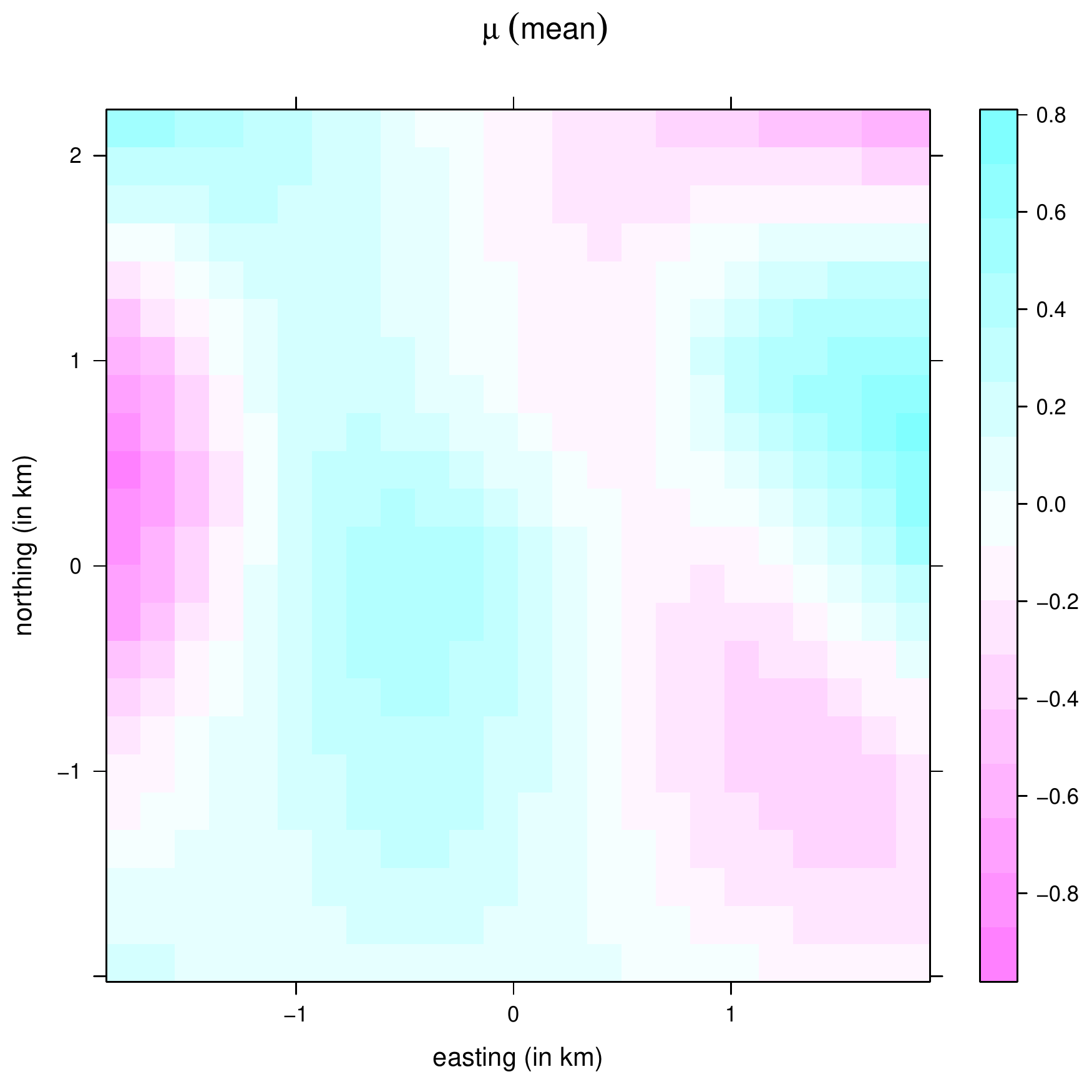}
\end{subfigure}
\begin{subfigure}{.5\textwidth}
\centering
\includegraphics[width=.8\linewidth]{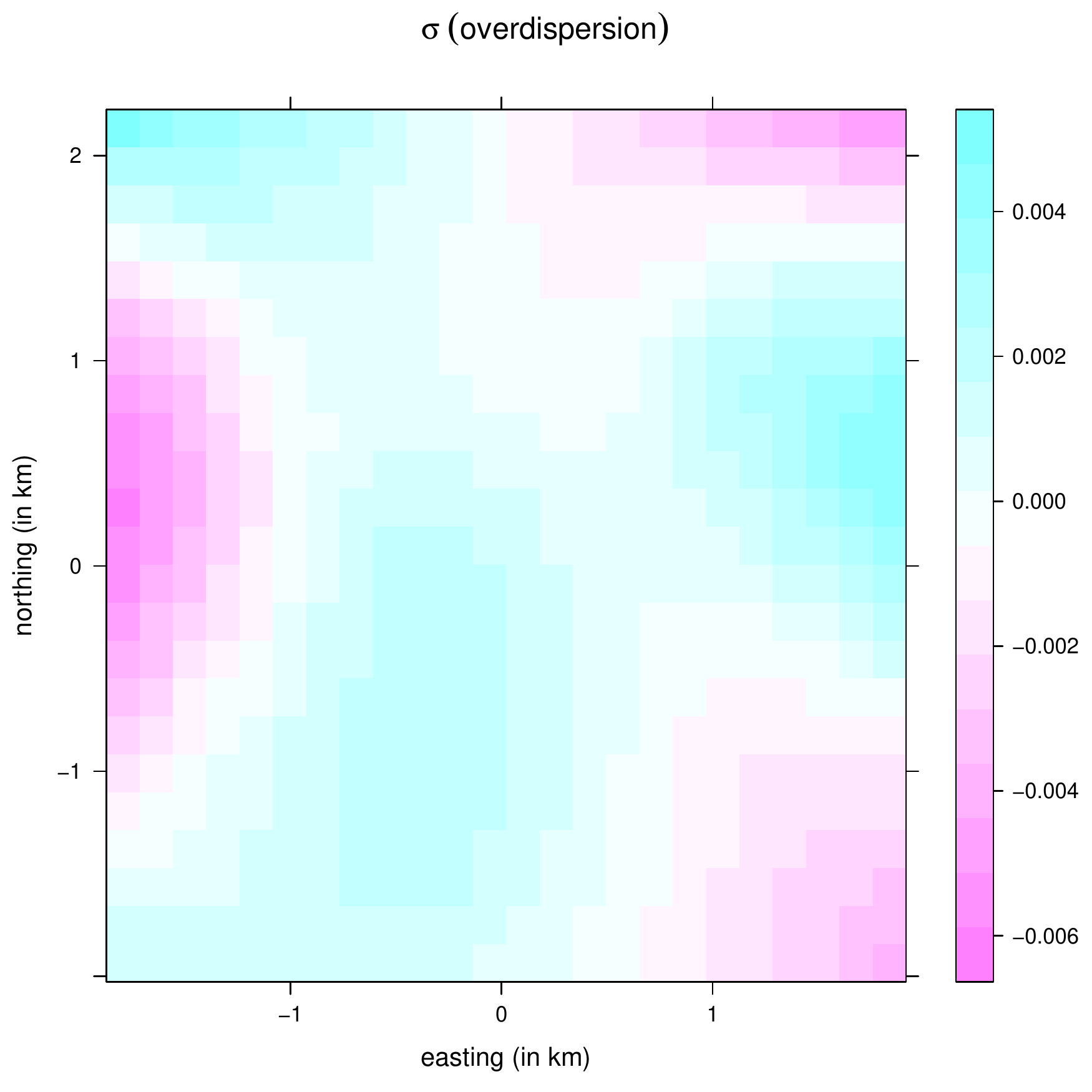}
\end{subfigure}
\caption{Spatial effects for mean (upper figure) and overdispersion (lower figure) of seabird population.}
\label{fig:spatial_eff}
\end{figure}

Subsampling yielded an optimal $m_\text{stop}$ of $2231$, split in $m_{\text{stop},\mu}=1871$ and $m_{\text{stop},\sigma}=336$. The resulting model selected 46 out of 48 possible covariates in $\mu$ and 8 out of 48 in $\sigma$, which is far too complex of a model (especially in $\mu$) to be useful. 

\begin{figure}
\centering
\includegraphics[width=0.45\textwidth]{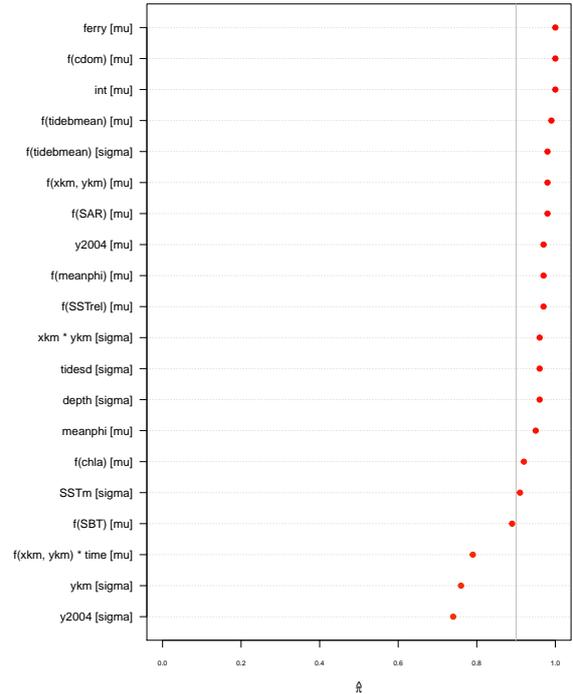}
\caption{Selection frequencies of the $20$ most frequently selected biophysical covariate base-learners of common eider abundance, determined by stability selection with $q=35$ and $\text{PFER} = 6$. The grey line represents the corresponding threshold of $0.9$}
\label{fig:stabs_seabird}
\end{figure}

With stability selection (see Figure~\ref{fig:stabs_seabird}), 10 effects were selected for the location: the intercept, relative sea surface temperature (smooth), chlorophyll-a levels (smooth), chromophoric dissolved organic material levels (smooth), sea floor sediment grain size (linear and smooth), sea floor surface area (smooth), mean epidenthic tidal velocity (smooth), a smooth spatial interaction, the presence of nearby ferry routes (yes/no), and two factors to account for changes in 2004 and 2005 compared to the the year 2003. For the overdispersion parameter 5 effects were selected: sea surface temperature (linear), bathymetry (linear), the mean (smooth) and standard deviation (linear) of the epibenthic tidal velocity, and the linear spatial interaction. For the location, all metric variables entered the model nonlinearly. Only sediment grain size was selected linearly as well as nonlinearly in the model. The converse was true for the overdispersion parameter: only the mean epibenthic velocity was selected as a smooth effect and all others were selected as linear effects. In Figure~\ref{fig:spatial_eff} the spatial effects for the mean and overdispersion can be seen.

\subsection{Comparison to results of the cyclic method}

Comparing the model with the results of Smith \emph{et al.}~\cite{smith2016seabirds}, the noncyclical model was larger in $\mu$ (10 effects, compared to 8 effects), but smaller in $\sigma$ (5 effects, compared to 7 effects). Chlorophyll-a levels, mean epibenthic tidal velocity, smooth spatial variation and year were not selected for the mean by stability selection with the cyclical fitting algorithm. On the other hand bathymetry was selected by the cyclical fitting method, but not by the noncyclical. For the overdispersion parameter the cyclical algorithm selected the year and the northing of a segment (the north-south position of a segment relative to the median) in addition to all effects selected by the noncyclical variant. Most effects were selected by both the cyclical and the noncyclical algorithm and the differences in the selected effects were rather small. 

In the simulation study for the negative binomial distribution (Section \ref{sec:simulation}), the noncyclical variant had a smaller false positive rate and a higher true positive rate. Even though the simulation was simplified compared to this application (only linear effects, known true number of informative covariates, uncorrelated effects), the results suggest to prefer the noncyclical variant. Nonetheless, the interpretation of selected covariate effects and final model assessment rests ultimately with subject matter experts.

\section{Conclusion}
\label{sec:conclusion}

The main contribution of this paper is a statistical model building algorithm that combines the three approaches of gradient boosting, GAMLSS and stability selection. As shown in our simulation studies and the application on sea duck abundance in Section \ref{sec:application}, the proposed algorithm incorporates the flexibility of structured additive regression modeling via GAMLSS, while it simultaneously allows for a data-driven generation of sparse models.

Being based on the gamboostLSS framework by Mayr \emph{et al.} \cite{mayr2012generalized}, the main feature of the new algorithm is a new ``noncyclical'' fitting method for boosted GAMLSS models. As shown in the simulation studies, this method does not only increase the flexibility of the variable selection mechanism used in gamboostLSS, but is also more time-efficient than the traditional cyclical fitting algorithm. In fact, even though the initial runtime to fit a single model may be higher (especially if the base-learner selection is done via the outer loss approach), this time is regained while finding the optimal number of boosting iterations via cross-validation approaches. Furthermore, the convergence speed of the new algorithm proved to be faster, and consequently fewer boosting iterations were needed in total.

Regarding stability selection, we observed that the noncyclical algorithm often had fewer false positives as well as more true positives compared to the cyclical variant in the two-parameter distributions tested in our simulation study. For high dimensional cases, however, the differences between both methods reduced and, especially with regard to the number of true positives, approximately equal results were achieved. For three-parameter distributions the cyclical variant achieved better values throughout with respect to both true and false positive rates. This may be due to the the fact that for more complex distributions, similar densities can be achieved with different parameter settings. For example, in a zero-inflated negative binomial setting, a small location may be hard to distinguish from a large zero-inflation. Obviously, the behavior of the cyclical variant is more robust in these situations than the noncyclical variant, which tends to fit very different models on each subsample and consequently selects a higher amount of non-informative variables. 

In summary, we have developed a framework for model building in GAMLSS that simplifies traditional optimization approaches to a great extent. For practitioners and applied statisticians, the main consequence of the new methodology is the incorporation of fewer noise variables in the GAMLSS model, leading to sparser and thus more interpretable models. Furthermore, the tuning of the new algorithm is far more efficient and leads to much shorter run times, particularly for complex distributions.

\section*{Implementation}
\label{sec:implementation}

The derived fitting methods for \texttt{gamboostLSS} models are implemented in the R
add-on package \textbf{gamboostLSS} \cite{hofner2016gamboostLSS}. The fitting algorithm can be specified 
via the \texttt{method} argument. By default \texttt{method} is set to \texttt{"cyclical"}, which is the originally proposed algorithm. Both new noncyclical algorithms can be selected via \texttt{method = "inner"} and \texttt{method = "outer"}. 
Base-learners and some of the basic methods are implemented in the R package
\textbf{mboost} \cite{mboost2010,Hofner:mboost:2014,mboost2016package}. 
The basic fitting algorithm for each distribution parameter is also implemented in \textbf{mboost}. 
For a tutorial and an explanation of technical details of \textbf{gamboostLSS} 
see \cite{gamboostLSS2015tut}. Stability selection is implemented in the R package \textbf{stabs} \cite{pkg:stabs:CRAN:0.5,hofner2014controlling}, 
with a specialized function for \texttt{gamboostLSS} models which is included
in \textbf{gamboostLSS} itself. The source code of \texttt{mboost}, \texttt{gamboostLSS} and \texttt{stabs} is hosted openly at
\begin{itemize}
\item[] \url{http://www.github.com/boost-R/mboost}
\item[] \url{http://www.github.com/boost-R/gamboostLSS}
\item[] \url{http://www.github.com/hofnerb/stabs}.
\end{itemize}

\section*{Acknowledgements}
\sloppy{We thank Mass Audubon for the use of common eider abundance data.}

\bibliographystyle{spmpsci}
\bibliography{bib.bib}

\end{document}